\documentclass[11pt]{article}

\ifx\Red\relax\def\Red#1{#1}\fi
\def\Red#1{#1}

\usepackage{epsfig}
\usepackage{amssymb}
\usepackage{verbatim}
\usepackage{delarray}
\usepackage{graphics}
\usepackage{epic}
\usepackage{stfloats}
\usepackage{amsmath}

\newcommand{\lra}{\leftrightarrow}
\newcommand{\inner}[2]{\langle{#1},{#2}\rangle}

\newcommand{\A}{{\mathcal{A}}}

\newcommand{\CC}{{\mathcal{C}}}

\newcommand{\FF}{{\mathcal{F}}}
\newcommand{\G}{{\mathcal{G}}}

\newcommand{\I}{{\mathcal{I}}}

\newcommand{\PP}{{\mathcal{P}}}

\newcommand{\RR}{{\mathcal{R}}}
\newcommand{\SSS}{{\mathcal{S}}}

\newcommand{\V}{{\mathcal{V}}}
\newcommand{\U}{{\mathcal{U}}}
\newcommand{\W}{{\mathcal{W}}}

\newcommand{\F}{{\mathbb{F}}}
\newcommand{\R}{{\mathbb{R}}}
\newcommand{\Z}{{\mathbb{Z}}}
\newcommand{\zerob}{{\mathbf 0}}

\newcommand{\ab}{{\mathbf a}}

\newcommand{\cb}{{\mathbf c}}

\newcommand{\gb}{{\mathbf g}}

\renewcommand{\sb}{{\mathbf s}}
\newcommand{\tb}{{\mathbf t}}

\newcommand{\Gb}{{\mathbf G}}
\newcommand{\Hb}{{\mathbf H}}

\newcommand{\Vb}{{\mathbf V}}

\newcommand{\Bf}{{\mathfrak{B}}}

\newcommand{\ie}{{\em i.e., }}
\newcommand{\eg}{{\em e.g., }}

\newcommand{\openbox}{\leavevmode
     \hbox to.77778em{%
     \hfil\vrule
     \vbox to.675em{\hrule width.6em\vfil\hrule}%
     \vrule\hfil}}
\newcommand{\qed}{\hspace*{1cm}\hspace*{\fill}\openbox}

\begin{document}

\title{Codes on Graphs:  Fundamentals}

\author{G. David Forney, Jr.\thanks{
Laboratory for Information and Decision Systems,
Massachusetts Institute of Technology,
Cambridge, MA 02139
 (email: forneyd@comcast.net).  Part of this paper was presented at the 2012 Allerton Conference \cite{F12}.}}
 \date{}

\maketitle

\begin{abstract}  This paper develops a fundamental theory of realizations of linear and group codes on general graphs using elementary group theory, including basic group duality theory.  
Principal new and extended results include: normal realization duality;  analysis of systems-theoretic properties of fragments of realizations and their connections;  ``minimal $\Leftrightarrow$ trim and proper" theorem for cycle-free codes;   results  showing that all constraint codes except interface nodes may be assumed to be trim and proper, and that the interesting part of a cyclic realization is its ``2-core;" notions of observability and controllability for fragments, and related tests; relations between state-trimness and controllability, and dual state-trimness and observability.
\end{abstract}


\section{Introduction}

The subject of ``codes on graphs" was founded by Tanner in \cite{T81}. Tanner showed how codes such as low-density parity-check (LDPC) codes may be defined by a graphical system of variables and constraints, and decoded by generic iterative decoding algorithms (``sum-product," ``max-sum") that are optimal in the cycle-free case.

After  years of obscurity, Tanner's results were rediscovered (largely independently) in Wiberg's  doctoral  thesis \cite{W96, WLK95}, which showed that capacity-approaching codes (\eg turbo codes and LDPC codes) and their decoding algorithms could be viewed within a common graphical framework.  By introducing internal (``state") variables, Wiberg  made connections with  topics such as  convolutional codes, trellis codes, and tail-biting trellis codes, as well as with classical linear systems theory.

The first paper \cite{F01} in this series on codes on graphs used an algebraic approach to realizations of linear and group codes by systems of variables and constraints, as in the behavioral systems theory of Willems \cite{W89}.  Its key observation was that, without any loss of generality or  essential increase of complexity, every realization may be assumed to be ``normal;" \ie to satisfy the ``normal degree constraints" that all external variables (``symbols")  have degree 1, whereas all internal variables (``states")  have degree 2.  Two important consequences are:
\begin{itemize}
\item (\textbf{Normal graph}) A normal realization is naturally represented by a ``normal graph."
\item (\textbf{Normal realization duality}) A linear or group normal realization $\RR$ has a well-defined dual realization $\RR^\circ$, such that   if $\RR$ realizes a linear or group code $\CC$, then $\RR^\circ$  realizes $\CC^\perp$.
\end{itemize}

Most current work on graph representations of codes uses the language of factor graphs \cite{KFL01}.  As is well known, there is a one-to-one correspondence between normal graphs of realizations and normal factor graphs of indicator functions of realizations \cite{L04}, as well as a nice generalization of normal realization duality  to normal factor graph duality \cite{AM11, F11b}.  However, for the purposes of this paper, we prefer to stay at the most basic level.

In this paper, we study the fundamental systems-theoretic properties of general normal linear and (finite abelian) group realizations, in some cases generalizing the results of \cite{FGL12} for linear tail-biting trellis realizations, and those of \cite{FT93, FT04} for conventional trellis realizations over groups.  
Our results apply not just to codes, but to any linear or group system defined by a network of variables and constraints;  \eg classical linear systems, or various types of physical systems.  However, we  use the language of coding theory.

As in Willems \cite{W89} and Vontobel and Loeliger \cite{VL03}, we analyze a realization by cutting it into fragments, or by combining fragments into larger fragments.  The smallest fragment is a single constraint;  the largest fragment is the complete realization.  

In Section \ref{Section 2}, we begin a systematic study of fragments of realizations, which are more complex than constraints or realizations in that they have both internal and external state variables.  

In Section \ref{Section 3}, we review elementary group theory, including elementary duality theory for finite abelian groups.  We rely on group theory rather than linearity throughout this paper, not because group codes are so important, but rather because in this setting we feel that  any result that  cannot be proved by elementary group theory is probably not fundamental.

In Section 4, we introduce extended behaviors, which yield  nice proofs of the normal realization duality theorem \cite{F01} and basic controllability/observability results for realizations \cite{FGL12}.  We also introduce a generalization of normal realization duality theorem that has greater symmetry between primal and dual domains.

Section \ref{Section 4} investigates the external properties of fragments, which generalize those of constraint codes.  We begin with the fundamental theorem for subdirect products (FTSP), which highlights the significance of what we call interface nodes in behavioral realizations.  Using the FTSP, we generalize the results of \cite{FGL12} on trimness, properness, and local reducibility of constraint codes to fragments, and show that we may assume that all constraints other than interface nodes are trim and proper.  Finally, we show that the external state space of a trim and proper leaf fragment is uniquely determined, up to isomorphism.

In Section 6, we first review some elementary graph theory.  Using the leaf fragment theorem, we give an improved derivation of the ``minimal $\Leftrightarrow$ trim + proper" theorem of \cite{FGL12}, which shows that every state space in a trim and proper cycle-free realization is uniquely determined, up to isomorphism.  We also show that a cyclic realization may be partitioned into a number of cycle-free leaf fragments, which act as static interface nodes, and a leafless cyclic ``2-core," which is the essential dynamical core of the realization.

Section \ref{Section 5} studies the observability and controllability properties of fragments, which resemble  those of realizations.  We consider several different notions of observability, and the dual notions of controllability.  We compare and contrast Willems'  \cite{W07} proposed notion of behavioral controllability for $n$-dimensional systems.  We establish many relations between these properties.  We give a different derivation of the result of \cite{FGL12} that in a trim and proper realization, any internal unobservability or uncontrollability must reside within its 2-core.  Finally, we derive relations between state-trimness and our notions of controllability, and the dual relations between dual state-trimness and our notions of observability.

\section{Realizations and fragments}\label{Section 2}

In this section, we review realizations, normal realizations, normal graphs, and  behaviors.  We  introduce extended behaviors, which, while redundant, yield nice proofs.  Finally, we formally introduce fragments. 

\subsection{Realizations}

In this paper, as in \cite{F01}, a \emph{realization} $\RR$ of a code $\CC$ will be defined  by a system of variables and constraints, in the style of behavioral systems theory \cite{W89}.  

We distinguish two types of variables.  \emph{External variables}, or \emph{symbols}, are the symbols of the code $\CC$ that is being realized.  \emph{Internal variables}, or \emph{states}, are additional auxiliary variables introduced by the designer of the realization for some purpose.  The key difference is that internal variables may be changed at will by the designer, whereas external variables are fixed \emph{a priori}.\footnote{In Section \ref{TC}, we will observe another, more graphical distinction: symbol variables must always lie on the boundary or outside of the cyclic ``2-core;"  \ie every  variable inside the 2-core is a state variable.}

A finite realization $\RR$ is  defined by a finite set $A(\RR)$ of \emph{symbol alphabets} $\A_k$, a finite set $S(\RR)$ of \emph{state alphabets} $\SSS_j$, and  a finite set $C(\RR)$ of \emph{constraint codes} $\CC_i$. 
We define the \emph{symbol configuration space} as the Cartesian product $\A = \Pi_{A(\RR)} \A_k$, and the \emph{state configuration space} as $\SSS = \Pi_{S(\RR)} \SSS_j$.  

For a \emph{linear} or \emph{group realization}, each \emph{variable} $V$  takes values $v$ in a finite-dimensional vector space $\V$ over a base field $\F$, or in a finite abelian group $\V$, respectively.  A Cartesian product of variable alphabets then becomes an (external) direct product of these vector spaces or groups.  

We will often identify a variable by its \emph{alphabet} $\V$, or by its \emph{value} $v \in \V$.  In the linear case, the ``size" of  $V$ will be measured by the dimension $\dim \V$ of its alphabet, whereas in the group case it will be measured by the size $|\V|$ of its alphabet.  Otherwise we will use common notation for linear and group realizations.

A  \emph{constraint code} (or simply \emph{constraint}) $\CC_i$  \emph{involves} subsets $A(\CC_i) \subseteq A(\RR)$ and $S(\CC_i) \subseteq S(\RR)$ of the symbol and state variables, respectively.  Thus $\CC_i \subseteq \A^{(i)} \times \SSS^{(i)}$, where $\A^{(i)} = \prod_{A(\CC_i)} \A_k$ and $\SSS^{(i)} = \prod_{S(\CC_i)} \SSS_j$. The elements $(\ab^{(i)}, \sb^{(i)})$ of $\CC_i$ are called \emph{valid configurations}.  In a linear or group realization, $\CC_i$ must be a subspace or subgroup of $\A^{(i)} \times \SSS^{(i)}$;  \ie a linear or group code.  

Some simple  constraints are:
\begin{itemize}
\item An \emph{equality constraint} is defined by a repetition code $\CC_{=\V}$ of length $n$ over a set of $n$ variables with a common alphabet $\V$.  In a graphical representation, an equality constraint will be denoted by a box containing an equals (=) sign.
\item A \emph{zero-sum constraint} is defined by a zero-sum (single-parity-check) code $\CC_{+\V}$ of length $n$ over a set of $n$ variables with a common linear or group alphabet $\V$.   In a graphical representation, a zero-sum constraint will be denoted by a box containing a plus (+) sign.
\item A \emph{sign-inversion constraint} is defined by a zero-sum code $\CC_{\sim\V}$ of length 2;  \ie the constraint is $v_1 + v_2 = 0$, or equivalently $v_1 = -v_2$.  In a graphical representation, a sign-inversion constraint will be denoted by a box containing a negation ($\sim$) sign, or  by a small circle ($\circ$) (a  convention borrowed from digital logic diagrams, introduced previously in \cite{MK05}).
\end{itemize}
 
 \pagebreak
 The \emph{(internal) behavior} $\Bf$ of  $\RR$ is the set of \emph{valid configurations} $(\ab, \sb) \in \A \times \SSS$ for which all constraints are satisfied;  \ie such that $(\ab^{(i)}, \sb^{(i)}) \in \CC_i$ for all $\CC_i \in C(\RR)$.   If $\RR$ is a linear or group realization, then $\Bf$ is a subspace or subgroup of $\A \times \SSS$. The \emph{code} or \emph{external behavior} $\CC$ realized by $\RR$ is the projection $\Bf_{|\A}$;  \ie the set of all symbol configurations $\ab \in \A$ such that $(\ab, \sb) \in \Bf$ for some $\sb \in \SSS$.  If $\RR$ is a linear or group realization, then $\CC$ is a subspace or subgroup of $\A$;  \ie a linear or group code.  Two realizations are \emph{equivalent} if they realize the same code $\CC$.
 
 \subsection{Normal realizations}
 
 We define the \emph{degree} of a variable as the number of constraints in which it is involved.  A \emph{realization} is \emph{normal} if all symbol variables have degree 1, and all nontrivial state variables have degree 2.\footnote{Degree-1 state variables do not impose any constraints on a realization, so their alphabets may be assumed to be trivial, or they may be deleted.  In this paper, we will assume that degree-1 state variables have been deleted.} 
 
 As shown in \cite{F01}, any realization may be ``normalized" with essentially no change in realization complexity by replacing variables by equality constraints between replica variables.
  In view of this simple conversion, we may and will assume that all realizations are normal.
  
 A normal realization is naturally represented by a \emph{normal graph} \cite{F01}, in which constraints are represented by vertices, state variables by edges, and symbol variables by half-edges, with an edge (resp.\ half-edge) incident on the vertices (resp.\ vertex) that represent(s) the constraints (resp.\ constraint) in which the corresponding variable is involved. 
 
 We say that a normal realization is connected if its graph is connected.
  If a normal graph is disconnected, then the code that it realizes is the Cartesian product of the codes realized by each component.  Therefore we may and will assume that all normal realizations are connected.
  
  \subsection{Fragments}

We now begin our formal study of  fragments of normal realizations.  A fragment may range from a single constraint code to the entire realization.  As we will see, realizations may be analyzed by studying  the effects of connecting or disconnecting fragments.

  If $\RR$ is a normal realization with a connected normal graph $\G$, then a \emph{fragment} $\FF$ of $\RR$ is a part of $\RR$ that corresponds to a connected subgraph $\G^\FF$ of $\G$ obtained by ``cutting" certain edges of $\G$ into two half-edges.  (Vontobel and Loeliger call this operation ``drawing a box" \cite{VL03}.)  
  
  Thus whereas a normal graph $\G$ has three kinds of elements, namely  constraint vertices, state edges, and symbol half-edges, a fragment has a fourth kind: a state half-edge.  We call the corresponding state variable an \emph{external state variable}, relative to the fragment $\FF$.  The set $\partial(\FF) \subseteq S(\RR)$ of   external state variables of $\FF$ will be called the \emph{boundary} of $\FF$.

A fragment $\FF$ thus contains a nonempty subset  $C(\FF) \subseteq C(\RR)$ of the constraint codes of $\RR$ (vertices of $\G$), and the corresponding subset $A(\FF) \subseteq A(\RR)$ of symbol variables of $\RR$ (half-edges of $\G$) that are involved in these constraint codes.  It further contains the subset $S(\FF) \subseteq S(\RR)$ of the state variables of $\RR$ (edges of $\G$) that are involved in two of the constraint codes of $\FF$ as \emph{internal state variables}, again relative to the fragment $\FF$, as well as the  set  $\partial(\FF) \subseteq S(\RR)$ of external state variables of $\FF$. 
The respective configuration spaces will be denoted by 
  $\A^{\FF} = \prod_{A(\FF)} \A_k$, $\SSS^{\FF,\mathrm{int}} = \prod_{S(\FF)} \SSS_j$, and $\SSS^{\FF,\mathrm{ext}} = \prod_{\partial(\FF)} \SSS_j.$
  
We note two important special cases.  A fragment with no internal state variables is  a single constraint code $\CC_i$.  A fragment with no external state variables is an entire normal realization $\RR$. 
  
  A fragment has two kinds of half-edges, corresponding to  symbol variables and external state variables, respectively.  To maintain this distinction, we will continue to represent symbol variables in figures by our usual ``dongle" symbol ($\vdash$), whereas we will represent  external state variables by an ordinary half-edge (--).  
    
  \vspace{1ex}
  \noindent
  \textbf{Example 1} (trellis fragment).
   For example, Figure \ref{TF} shows a fragment $\FF^{[j,k)}$ of a trellis realization $\RR$, as in \cite{GLF12}.  Its constraint codes are $C(\FF^{[j,k)}) = \{\CC_i : i \in [j,k)\}$; its symbol variables are $A(\FF^{[j,k)}) = \{\A_i : i \in [j,k)\}$; its internal state variables are $S(\FF^{[j,k)}) = \{\SSS_i : i \in (j,k)\}$; and its external state variables are $\partial(\FF^{[j,k)}) = \{\SSS_j, \SSS_k\}$. \qed

\begin{figure}[h]
\setlength{\unitlength}{5pt}
\centering
\begin{picture}(70,9)(-2, 3)
\put(25,2.5){\framebox(5,5){$\CC_{j+1}$}}
\put(55,2.5){\framebox(5,5){$\CC_{k-1}$}}
\multiput(-2,5)(50,0){2}{\line(1,0){7}}
\multiput(30,5)(30,0){2}{\line(1,0){7}}
\multiput(10,5)(20,0){1}{\line(1,0){15}}
\multiput(5,2.5)(40,0){1}{\framebox(5,5){$\CC_j$}}
\multiput(7.5,7.5)(20,0){2}{\line(0,1){3}}
\multiput(6,10.5)(20,0){2}{\line(1,0){3}}
\put(6,11){$\A_j$}
\put(26,11){$\A_{j+1}$}
\put(42,5){$\cdots$}
\multiput(57.5,7.5)(20,0){1}{\line(0,1){3}}
\multiput(56,10.5)(20,0){1}{\line(1,0){3}}
\put(56,11){$\A_{k-1}$}
\put(0,6){$\SSS_{j}$}
\put(15,6){$\SSS_{j+1}$}
\put(32,6){$\SSS_{j+2}$}
\put(63,6){$\SSS_k$}
\put(49,6){$\SSS_{k-1}$}
\end{picture}
\caption{Fragment $\FF^{[j,k)}$ of a trellis realization $\RR$.}
\label{TF}
\end{figure}

  The \emph{internal behavior} $\Bf^\FF$ of a fragment $\FF$ is the set of all configurations $(\ab^{\FF}, \sb^{\FF,\mathrm{ext}}, \sb^{\FF,\mathrm{int}}) \in \A^{\FF} \times \SSS^{\FF,\mathrm{ext}} \times\SSS^{\FF,\mathrm{int}}$ that satisfy all constraints $\CC_i$ for all $\CC_i \in C(\FF)$. Its \emph{external behavior} $\CC^{\FF}$ is the projection of $\Bf^{\FF}$ on $\A^{\FF} \times \SSS^{\FF,\mathrm{ext}}$.    
    If $\FF$ is an entire normal realization $\RR$, then it has no external state variables,  and these definitions reduce to those for a normal realization.  If $\FF$ is a constraint code $\CC_i$ with no internal state variables, then $\Bf^{\FF} = \CC^{\FF} = \CC_i$.
  
Notice that if we conflate  symbol and external state half-edges, then a fragment $\FF$ is a normal realization of its external behavior $\CC^{\FF}$.  
  
  The \emph{degree} $\deg(\FF)$ of a fragment $\FF$ will be defined as the number of its external state variables; \ie the size $|\partial(\FF)|$ of its boundary.   As already noted, a fragment of degree 0 is simply a normal realization.  Fragments of degrees 1, 2, 3, and $\ge 4$ will be called \emph{leaf, trellis, cubic} and \emph{hypercubic} fragments, respectively.
  
\section{Groups, vector spaces and duality}\label{Section 3}

In this section, we will develop in parallel the general principles of realizations of group codes over finite abelian groups and of linear codes over a field $\F$.  The two theories involve similar algebra, and indeed coincide when $\F$ is a finite field.  All proofs will be group-theoretic, and therefore we will generally use group-theoretic language, although from time to time we will translate our results into vector space (linear algebra) terminology, which is no doubt more familiar to most readers.
  
  \subsection{Finite abelian groups and vector spaces}

If $G$ is an abelian group, then any subgroup $H \subseteq G$ is automatically normal\footnote{There is no connection between ``normal subgroups" and ``normal realizations."}, and the cosets of $H$ in $G$ form a quotient group $G/H$.  If $G$ is finite, then $|G| = |H||G/H|$, and given any set $[G/H]$ of coset representatives for $G/H$, every element $g \in G$ may be uniquely represented as a sum $g = h + r$ with $h \in H, r \in [G/H]$.  However, in general $[G/H]$ cannot be taken as a subgroup of $G$;  \eg there is no subgroup of $\Z_4$ that can be taken as a set of coset representatives for $\Z_4/2\Z_4$.

\pagebreak
More generally, a \emph{normal series} is a chain of normal subgroups $G_n = \{0\} \subseteq G_{n - 1} \subseteq \cdots \subseteq G_0 = G$.  The \emph{factor groups} of the normal series are the quotient groups $G_i/G_{i+1}$.  (Alternatively, we may let $G_n \neq \{0\}$, in which case $G_n$ will be regarded as a factor group.)   If $G$ is finite, then we have $|G| = \prod_{i} |G_i/G_{i+1}|$.  If $[G_i/G_{i+1}], 0 \le i \le n-1$, are any sets of coset representatives for each quotient group, then the elements of $G$ may be uniquely represented as sums of coset representatives, one from each set;  this is called a \emph{chain coset decomposition}.

Similarly, if $V$ is a finite-dimensional vector space over $\F$, then  the cosets of any subspace $W$ in $V$ form a quotient space $V/W$ such that $\dim V = \dim W + \dim V/W$.  For vector spaces, it is always possible to find a coset representative subspace $[V/W] \subseteq V$ of dimension $\dim V/W$ such that $V$ is equal to the (internal) direct product $W \times [V/W]$, so any basis for $W$ and basis for $[V/W]$ together form a basis for $V$.

More generally, a normal series of subspaces of a vector space $V$ is a chain of subspaces $V_n = \{0\} \subseteq V_{n - 1} \subseteq \cdots \subseteq V_0 = V$.  The factors of the normal series are the quotient spaces $V_i/V_{i+1}$.  (If we let $V_n \neq \{0\}$, then $V_n$ is also a factor.)   If $V$ is finite-dimensional, then we have $\dim V = \sum_{i} \dim V_i/V_{i+1}$.  If $[V_i/V_{i+1}], 0 \le i \le n-1$, are coset representative subspaces for the quotient spaces, then $V$ is the direct product of the subspaces $[V_i/V_{i+1}], 0 \le i \le n-1$, so the union of any set of bases for these subspaces is a basis for $V$.  Thus in the vector space case, unlike the group case, the order of  factors in a normal series does not matter.

The \emph{fundamental theorem of homomorphisms} says that if $f: G \to H$ is a homomorphism with image $f(G)$ and kernel $K \subseteq G$, then $K$ is a normal subgroup of $G$, and $f(G) \cong G/K$.  On the other hand, if $H$ is any normal subgroup of $G$, then the \emph{natural map}  $\pi: G \to G/H$ defined by $\pi(g) = H + g$ is a homomorphism with kernel $H$ and image $G/H$.

The \emph{correspondence theorem} says that if $H$ is a normal subgroup of $G$, and $f$ is a homomorphism such that $f: G \to f(G)$ and $f: H \to f(H)$ have the same kernel $K \subseteq H \subseteq G$, then $G/H \cong f(G)/f(H)$.  Thus the natural map $\pi: G \to G/G_n$ maps a normal series $G_n \subseteq G_{n - 1} \subseteq \cdots \subseteq G_0 = G$ to a normal series of quotients, $\{0\} \subseteq G_{n - 1}/G_n \subseteq \cdots \subseteq G/G_n$, whose factor groups are isomorphic to those of the original series, apart from $G_n$.

\subsection{Duality}

A finite abelian group $G$ has a \emph{dual group} $\hat{G}$ (namely, its character group) such that there exists a well-defined \emph{pairing} $\inner{\hat{g}}{g} \in \R/\Z$ for all $g \in G, \hat{g} \in \hat{G}$ that is bihomomorphic:  \ie $\inner{0}{g} = \inner{\hat{g}}{0} = 0$,  $\inner{\hat{g}_1 + \hat{g}_2}{g} = \inner{\hat{g}_1}{g} + \inner{\hat{g}_2}{g}$, and so forth.   
In the finite abelian case, $G$ and $\hat{G}$ are isomorphic.

If $H$ is a subgroup of $G$, then its \emph{orthogonal subgroup} is defined as $H^\perp = \{\hat{g} \in \hat{G} : \inner{\hat{g}}{h} = 0 \mathrm{~for~all~} h \in H\}$.  $H^\perp$ is a subgroup of $\hat{G}$, and $(H^\perp)^\perp = H$.  The product $|H||H^\perp|$ is equal to $|G| = |\hat{G}|$.  Moreover, $H^\perp$ acts as the dual group to the quotient group $G/H$, with the pairing $\inner{\hat{g}}{H + g} = \inner{\hat{g}}{g}$ for $\hat{g} \in H^\perp, H + g \in G/H$, so in the finite abelian case $H^\perp$ is actually isomorphic to $G/H$.  More generally, we have \emph{quotient group duality}:  if $J \subseteq H \subseteq G$, then the quotient group $J^\perp/H^\perp$ acts as the dual group to $H/J$. 

Similarly, a finite-dimensional vector space $V$ over $\F$ has a \emph{dual space} $\hat{V}$ of the same dimension such that there exists a well-defined \emph{inner product} $\inner{\hat{v}}{v} \in \F$ for all $v \in V, \hat{v} \in \hat{V}$ that is bilinear:  \ie $\inner{0}{v} = \inner{\hat{v}}{0} = 0$,  $\inner{\hat{v}_1 + \hat{v}_2}{v} = \inner{\hat{v}_1}{v} + \inner{\hat{v}_2}{v}$, and so forth.

If $W$ is a subspace of $V$, then the orthogonal subspace $W^\perp$  is a subspace of $\hat{V}$, and $(W^\perp)^\perp = W$.  The sum $\dim W + \dim W^\perp$ is equal to $\dim V = \dim \hat{V}$.  Moreover, $W^\perp$ acts as the dual space to $V/W$, with the inner product $\inner{\hat{v}}{W + v} = \inner{\hat{v}}{v}$ for $\hat{v} \in W^\perp, W + v \in V/W$.

If $\Gb = \Pi_k G_k$ is an  external direct product of a finite collection of groups or vector spaces $G_k$, then the dual group or space to $\Gb$ is $\hat{\Gb} = \Pi_k \hat{G}_k$, and the pairing or inner product between $\gb \in \Gb$ and $\hat{\gb} \in \hat{\Gb}$ is given by the componentwise sum
  $
  \inner{\hat{\gb}}{\gb} = \sum_k \inner{\hat{g}_k}{g_k}.
  $
 If $\Hb = \Pi_k H_k \subseteq \Vb$ is a direct product of subgroups or subspaces $H_k \subseteq G_k$, then the orthogonal group or space is the direct product $\Hb^\perp = \Pi_k (H_k)^\perp \subseteq \hat{\Gb}$. 
 
 \subsection{Projection/cross-section duality}
 
 The most useful duality relationship for us will be \emph{projection/cross-section duality}. Let $C$ be a \emph{subdirect product} \cite{Hall}--- \ie a subgroup (or subspace) of an external direct product $A \times B$, where $A$ and $B$ are groups (or vector spaces). Then the \emph{projection} of $C$ on $A$ is defined as $C_{|A} = \{a \in A :  (a,b) \in C \mathrm{~for~some~} b \in B\},$ and the \emph{cross-section} of $C$ on $A$ is defined as $C_{:A} = \{a \in A :  (a,0) \in C\}$ (following the notation of \cite{MK05}).  In general, $\{0\} \subseteq C_{:A} \subseteq C_{|A} \subseteq A$ is a normal series.
 
 The projection/cross-section duality theorem says that if $C^\perp \subseteq \hat{A} \times \hat{B}$ is the orthogonal subgroup (or subspace) to $C$, then $(C_{:A})^\perp = (C^\perp)_{|\hat{A}}.$ For completeness, we repeat the one-line proof of \cite{F01}:
 $$ a \in C_{:A} \Leftrightarrow (a,0) \in C \Leftrightarrow (a,0) \perp C^\perp \Leftrightarrow a \perp (C^\perp)_{|\hat{A}}. $$
 
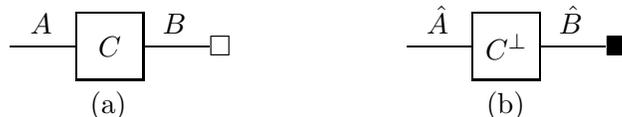
\begin{figure}[h]
\setlength{\unitlength}{5pt}
\centering
\begin{picture}(55,6)(-2, 2)
\put(0,5){\line(1,0){5}}
\multiput(5,2.5)(40,0){1}{\framebox(5,5){$C$}}
\put(10,5){\line(1,0){5}}
\put(1.5,6){$A$}
\put(11.5,6){$B$}
\put(15,4.3){$\square$}
\put(6,0){(a)}
\put(30,5){\line(1,0){5}}
\multiput(35,2.5)(40,0){1}{\framebox(5,5){$C^\perp$}}
\put(40,5){\line(1,0){5}}
\put(31.5,6){$\hat{A}$}
\put(41.5,6){$\hat{B}$}
\put(45,4.3){$\blacksquare$}
\put(36,0){(b)}
\end{picture}
\caption{(a) Cross-section $C_{:A}$;  (b)  projection $(C^\perp)_{|\hat{A}}$.}
\label{PCSD}
\end{figure}

We illustrate projection/cross-section duality in Figure \ref{PCSD}.  In Figure \ref{PCSD}(a), the constraint $C$ constrains the two variables $A$ and $B$.  We introduce a second constraint on $B$, namely the  zero  (``fixed," ``pinned," ``grounded") degree-1 constraint $C_\square = \{0\} \subseteq B$,  which we will  represent in figures by a special open-square symbol ($\square$).
Then $B$ becomes an internal state variable, and the resulting fragment can be seen to realize the cross-section $C_{:A} = \{a \in A : (a, 0) \in C\}$.

Similarly, in Figure \ref{PCSD}(b), the dual constraint, defined by the orthogonal code $C^\perp$, constrains the two dual variables $\hat{A}$ and $\hat{B}$.  We introduce a second constraint on $\hat{B}$, namely the  dummy  (``free," ``open") degree-1 constraint $\hat{C}_\blacksquare = \hat{B}$, which we will  represent in figures by a special closed-square symbol ($\blacksquare$).  Then $\hat{B}$ becomes an internal state variable, and the resulting fragment  can be seen to realize the projection $(C^\perp)_{|\hat{A}} = \{\hat{a }\in \hat{A} : (\hat{a}, \hat{b}) \in C^\perp  \mathrm{~for~some~} \hat{b} \in \hat{B}\}$.  

After normal realization duality has been discussed in Section \ref{NRDS}, it will become clear that Figure \ref{PCSD}(b) is the dual fragment to the fragment of Figure \ref{PCSD}(a), since $ (C_\square)^\perp = \{0\}^\perp = \hat{B} = \hat{C}_\blacksquare$.   (The sign inversions that usually appear in dual realizations are not needed, because for any abelian group $G$ we have $-G = G$.)

\subsection{Sum/intersection duality}
 
 Another useful duality relationship is \emph{sum/intersection duality}: if $A$ and $B$ are subgroups of $G$, then $(A + B)^\perp = A^\perp \cap B^\perp$, a subgroup of the dual group $\hat{G}$.  (Proof:  obviously $A + B \subseteq (A^\perp \cap B^\perp)^\perp$ and $(A + B)^\perp \subseteq A^\perp \cap B^\perp$;  together these relations imply $(A + B)^\perp = A^\perp \cap B^\perp$.)
 
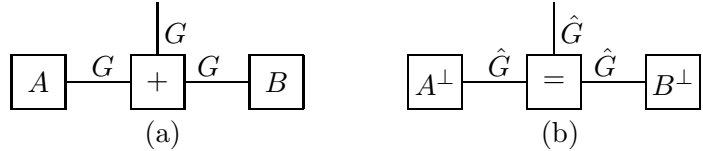
\begin{figure}[h]
\setlength{\unitlength}{5pt}
\centering
\begin{picture}(55,9)(-4, 2)
\multiput(-4,3)(40,0){1}{\framebox(4,4){$A$}}
\put(0,5){\line(1,0){5}}
\put(2,5.5){$G$}
\multiput(5,3)(40,0){1}{\framebox(4,4){$+$}}
\put(9,5){\line(1,0){5}}
\put(10,5.5){$G$}
\multiput(14,3)(40,0){1}{\framebox(4,4){$B$}}
\put(7,7){\line(0,1){4}}
\put(7.5,8){$G$}
\put(6,0.5){(a)}
\multiput(26,3)(40,0){1}{\framebox(4,4){$A^\perp$}}
\put(30,5){\line(1,0){5}}
\put(32,5.5){$\hat{G}$}
\multiput(35,3)(40,0){1}{\framebox(4,4){$=$}}
\put(39,5){\line(1,0){5}}
\put(40,5.5){$\hat{G}$}
\multiput(44,3)(40,0){1}{\framebox(4,4){$B^\perp$}}
\put(37,7){\line(0,1){4}}
\put(37.5,8){$\hat{G}$}
\put(36,0.5){(b)}
\end{picture}
\caption{(a) sum $A + B$;  (b)  intersection $A^\perp \cap B^\perp$.}
\label{SID}
\end{figure}

 \pagebreak
 We illustrate sum/intersection duality in Figure \ref{SID}.  In Figure \ref{SID}(a), the sum $A + B$ is realized by a symbol variable incident on a degree-3 zero-sum constraint $C_{+G}$, with the two other incident state variables  constrained to $A \subseteq G$ and $B \subseteq G$, respectively.  In Figure \ref{SID}(b), the intersection $A^\perp \cap B^\perp$ is realized by a dual symbol variable incident on a degree-3 equality constraint  $C_{=\hat{G}}$, with the two other  incident state variables  constrained to $A^\perp \subseteq \hat{G}$ and $B^\perp \subseteq \hat{G}$, respectively.
(Again, after Section \ref{NRDS}, it will become clear that Figures \ref{SID}(a) and \ref{SID}(b) depict dual fragments, since  equality and zero-sum constraints are duals.)

The sum $A + B$ is said to be an \emph{internal direct product} $A \times B$, and $A$ and $B$ are said to be \emph{independent}, if and only if $A \cap B = \{0\}$.  Then, and only then, every element  of $A + B$ may be expressed uniquely as a sum $a + b$, so $|A + B| = |A||B|$ in the group case, or $\dim (A + B) = \dim A + \dim B$ in the linear case.   In particular, $a +b = 0$ implies $a = b = 0$.  By sum/intersection duality, $A$ and $B$ are independent if and only if $A^\perp + B^\perp = \hat{G}$.

\subsection{Isomorphisms and adjoint isomorphisms}\label{AIS}

An \emph{isomorphism constraint} is specified by a degree-2 group code $C = \{(a,\varphi(a)) \in A \times B : a \in A\}$, where $A$ and $B$ are isomorphic groups and $\varphi:  A \to B$ is an isomorphism between them.  (In group theory, $C$ is called the \emph{graph} of the isomorphism $\varphi$.)  Degree-2 equality constraints  and sign-inversion constraints are examples of isomorphism constraints. 

The \emph{adjoint isomorphism} to $\varphi$ is the unique isomorphism $\hat{\varphi}:  \hat{B} \to \hat{A}$  such that $\inner{\hat{\varphi}(\hat{b})}{a} = \inner{\hat{b}}{\varphi(a)}$ for all $a \in \A, \hat{b} \in \hat{B}$.  In other words, $\hat{\varphi}$ is the unique isomorphism such that $\{(-\hat{\varphi}(\hat{b}), \hat{b})  : \hat{b} \in \hat{B}\}$ is the orthogonal code $C^\perp \subseteq  \hat{A} \times \hat{B}$ to $C$.
  For example, if $B = A$ and $\varphi_j$ is the equality isomorphism defined by $\varphi(a) = a$, then $\hat{B} = \hat{A}$, and the  adjoint isomorphism $\hat{\varphi}$ is  the equality isomorphism defined by $\hat{\varphi}(\hat{b}) = \hat{b}$.
  
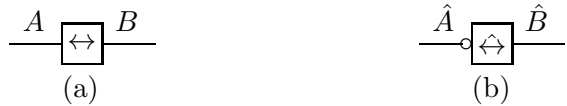
\begin{figure}[h]
\setlength{\unitlength}{5pt}
\centering
\begin{picture}(40,6)(0, 2)
\multiput(1,5)(50,0){1}{\line(1,0){4}}
\put(2,6){$A$}
\multiput(5,3.5)(40,0){1}{\framebox(3,3){$\leftrightarrow$}}
\multiput(8,5)(50,0){1}{\line(1,0){4}}
\put(9,6){$B$}
\put(5,1){(a)}
\multiput(32,5)(50,0){1}{\line(1,0){3.3}}
\put(33,6){$\hat{A}$}
\put(35,4.5){$\circ$}
\multiput(36,3.5)(40,0){1}{\framebox(3,3){$\hat{\leftrightarrow}$}}
\multiput(39,5)(50,0){1}{\line(1,0){4}}
\put(40,6){$\hat{B}$}
\put(36,1){(b)}
\end{picture}
\caption{Dual isomorphism constraints $C$ and $C^\perp$ realized with negative adjoint isomorphisms.}
\label{AI}
\end{figure}

   In a graphical representation, an isomorphism constraint $C$ will be represented as a degree-2 constraint labelled by  a left-right arrow ($\lra$) sign, as in Figure \ref{AI}(a).  The orthogonal code $C^\perp$ then specifies the negative adjoint isomorphism constraint $-\hat{\varphi}$, which is  represented as a degree-2 constraint labelled by $\hat{\lra}$, plus a small circle representing a sign inversion as in Figure \ref{AI}(b).  (Again, after Section \ref{NRDS}, it will be clear that Figures \ref{AI}(a) and \ref{AI}(b) are duals.)

More generally, given any group homomorphism $\varphi:  A \to B$, its \emph{adjoint homomorphism} is defined as the unique homomorphism $\hat{\varphi}:  \hat{B} \to \hat{A}$  such that $\inner{\hat{\varphi}(\hat{b})}{a} = \inner{\hat{b}}{\varphi(a)}$ for all $a \in \A, \hat{b} \in \hat{B}$;  \ie such that  $C = \{(a,\varphi(a)) : a \in A\}$ and $C^\perp = \{(-\hat{\varphi}(\hat{b}), \hat{b})  : \hat{b} \in \hat{B}\}$ are orthogonal group codes.

\section{Normal realization duality}

  In this section, we  introduce the extended behavior  $\bar{\Bf} \subseteq \A \times \SSS \times \SSS$ of a normal realization.  Although the extended behavior $\bar{\Bf}$ is a redundant version of the behavior $\Bf \subseteq \A \times \SSS$, it leads to the simplest  proof that we know of the normal realization duality theorem, and also a nice development of the controllability properties of linear or group realizations.  We  also give a generalized normal realization duality theorem that yields greater symmetry between  primal and dual realizations.
    
    \vspace{-1ex}
  \subsection{Extended behaviors}
  
Again, a normal realization $\RR$ is defined by a symbol alphabet set $A(\RR)$, a state alphabet set $S(\RR)$, and a constraint code set $C(\RR)$.  By the normal degree restrictions, as we run through all constraint codes $\CC_i \in C(\RR)$, each symbol variable appears precisely once, and each nontrivial state variable precisely twice.  We  denote the two values of a given state variable $\SSS_j \in S(\RR)$ by $s_j$ and $s'_j$;  it does not matter which one  is primed.  Then each element of the  external direct product   $\U = \prod_{C(\RR)} \CC_i$ of all constraint codes is a configuration $(\ab, \sb, \sb') \in \A \times \SSS \times \SSS$, where $\A = \prod_{A(\RR)} \A_k$ and $\SSS = \prod_{S(\RR)} \SSS_j$.  We call $\U$ the \emph{extended configuration universe}.

We then define the \emph{extended behavior} $\bar{\Bf}$ as the set  of all valid configurations in $\U$, where the validity constraint is $\sb = \sb'$;  \ie $\bar{\Bf} = \{(\ab, \sb, \sb) \in \U\}$.  Evidently $\bar{\Bf}$ is isomorphic to the behavior $\Bf$ via projection onto $\A \times \SSS$, and the code $\CC$ realized by $\RR$ is the projection of $\bar{\Bf}$ or $\Bf$ onto $\A$.

We note that the extended behavior $\bar{\Bf}$ may be expressed as $\bar{\Bf} = \U \cap \V$, where $\V$ denotes the \emph{validity space} $\V = \{(\ab, \sb, \sb) \in \A \times \SSS \times \SSS\}$.  In other words, we have two sets of constraints, namely the code constraints of $\U$ and the equality constraints of $\V$, and  the valid configurations in $\A \times \SSS \times \SSS$ are precisely those that satisfy both sets of constraints.

\vspace{-1ex}
         \subsection{Normal realization duality}\label{NRDS}
         
We now define the dual realization to a normal linear or group realization $\RR$ as in \cite{F01}, and give the simplest proof we know of the normal realization duality theorem, following Koetter \cite{K02, FGL12}.

Given a normal realization $\RR$, its \emph{dual realization} $\RR^\circ$ is the normal realization defined by 
\begin{itemize}
\item the dual symbol alphabet set $\hat{A}(\RR^\circ)$, whose elements are the dual symbol alphabets $\hat{\A}_k$;
\item the dual state alphabet set $\hat{S}(\RR^\circ)$, whose elements are the dual state alphabets $\hat{\SSS}_j$;
\item the dual constraint code set $\hat{C}(\RR^\circ)$, whose elements are  the orthogonal constraint codes $(\CC_i)^\perp$;
\item finally, a set of sign inversion constraints $C_{\sim \hat{\SSS}_j}$ that impose the validity constraints $\hat{s}_j = -\hat{s}'_j$ on the two values of each  dual state variable $\hat{\SSS}_j \in  \hat{S}(\RR^\circ)$, in place of  the equality validity constraints $C_{= {\SSS}_j}$ of the primal realization $\RR$.
\end{itemize}

The \emph{dual extended configuration universe} is the external direct product $\prod_{\hat{C} \in \RR^\circ} (\CC_i)^\perp = \U^\perp$ of all dual constraint codes.  The \emph{dual extended behavior} is the set $\bar{\Bf}^\circ = \{(\hat{\ab}, \hat{\sb}, -\hat{\sb}) \in \U^\perp\}$  of all valid configurations $(\hat{\ab}, \hat{\sb}, -\hat{\sb}) \in \U^\perp$, where the dual validity constraint is $\hat{\sb} = -\hat{\sb}'$. The \emph{dual behavior} is $\Bf^\circ = (\bar{\Bf}^\circ)_{|\hat{\A} \times \hat{\SSS}} \cong \bar{\Bf}^\circ$, and  the \emph{dual code} that is realized by $\RR^\circ$ is $\CC^\circ = (\Bf^\circ)_{|\hat{\A}} = (\bar{\Bf}^\circ)_{|\hat{\A}}$.

We  define the \emph{check space} of $\RR$ as the orthogonal space $\bar{\Bf}^\perp$ to its extended behavior $\bar{\Bf}$.  Since $\bar{\Bf} = \U \cap \V$, 
 the check space may  be expressed as $\bar{\Bf}^\perp = \U^\perp + \V^\perp$, by sum/intersection duality, where the \emph{validity check space} $\V^\perp = \{(\zerob, \hat{\sb}, -\hat{\sb}) : \hat{\sb} \in \hat{\SSS}\}$ is the orthogonal space to $\V$. 
 
\noindent
\textbf{Lemma}.  The code $\CC^\circ$ realized by $\RR^\circ$ is the check space cross-section $(\bar{\Bf}^\perp)_{:\hat{\A}}$.

\vspace{1ex}
\noindent
\textsc{Proof}:  
For $(\hat{\ab}, \hat{\sb}, \hat{\sb}') \in \U^\perp$, the coset $(\hat{\ab}, \hat{\sb}, \hat{\sb}') + \V^\perp$ contains the element $(\hat{\ab}, \zerob, \zerob)$ if and only if $\hat{\sb} = -\hat{\sb}'$.  Thus $\CC^\circ = (\U^\perp + \V^\perp)_{:\hat{\A}}$.  \qed 

\vspace{1ex}
\noindent
\textbf{Theorem} (\textbf{normal realization duality}).  $\CC^\circ = \CC^\perp$.

\vspace{1ex}
\noindent
\textsc{Proof}:  
By projection/cross-section duality, 
 $\CC^\circ = (\bar{\Bf}^\perp)_{:\hat{\A}} = (\bar{\Bf}_{|\A})^\perp = \CC^\perp.$  \qed \vspace{1ex}

\begin{figure}[h]
\setlength{\unitlength}{5pt}
\centering
\begin{picture}(65,46)(-2, 2)
\put(0,45){\line(1,0){5}}
\put(0,43.5){\line(0,1){3}}
\put(1.5,46){$\A$}
\multiput(5,42.5)(40,0){1}{\framebox(5,5){$\U$}}
\put(10,43){\line(1,0){9}}
\put(10,47){\line(1,0){9}}
\put(11,47.5){$\sb \in \SSS$}
\put(19,47){\line(0,-1){1}}
\put(11,43.5){$\sb' \in \SSS$}
\put(19,43){\line(0,1){1}}
\multiput(18,44)(40,0){1}{\framebox(2,2){$=$}}
\put(6,40){(a)}
\put(0,35){\line(1,0){5}}
\put(0,33.5){\line(0,1){3}}
\put(1.5,36){$\A$}
\multiput(5,32.5)(40,0){1}{\framebox(5,5){$\U$}}
\put(10,33){\line(1,0){9}}
\put(10,37){\line(1,0){9}}
\put(11,37.5){$\sb \in \SSS$}
\put(19,37){\line(0,-1){1}}
\put(11,33.5){$\sb' \in \SSS$}
\put(19,33){\line(0,1){1}}
\multiput(18,34)(40,0){1}{\framebox(2,2){$=$}}
\put(20,35){\line(1,0){7}}
\put(21,35.5){$\tb \in \SSS$}
\put(27,34.3){$\blacksquare$}
\put(6,30){(b)}
\put(0,25){\line(1,0){5}}
\put(0,23.5){\line(0,1){3}}
\put(1.5,26){$\A$}
\multiput(5,22.5)(40,0){1}{\framebox(5,5){$\U$}}
\put(10,23){\line(1,0){9}}
\put(10,27){\line(1,0){9}}
\put(11,27.5){$\sb \in \SSS$}
\put(18.5,26){$\circ$}
\put(11,23.5){$\sb' \in \SSS$}
\put(19,23){\line(0,1){1}}
\multiput(18,24)(40,0){1}{\framebox(2,2){$+$}}
\put(20,25){\line(1,0){7}}
\put(21,25.5){$\tb \in \SSS$}
\put(27,24.3){$\square$}
\put(6,20){(c)}

\put(0,15){\line(1,0){5}}
\put(0,13.5){\line(0,1){3}}
\put(1.5,16){$\A$}
\multiput(5,12.5)(40,0){1}{\framebox(5,5){$\U$}}
\put(10,13){\line(1,0){8}}
\put(10,17){\line(1,0){8}}
\put(11,17.5){$\sb \in \SSS$}
\put(11,13.5){$\sb' \in \SSS$}
\multiput(18,16)(40,0){1}{\framebox(2,2){$=$}}
\put(19,16){\line(0,-1){2}}
\put(20,17){\line(1,0){7}}
\put(20.5,14.5){$\tb \in \SSS$}
\put(27,16.3){$\blacksquare$}
\multiput(18,12)(40,0){1}{\framebox(2,2){$=$}}
\put(20,13){\line(1,0){7}}
\put(27,12.3){$\blacksquare$}
\put(6,10){(d)}

\put(0,5){\line(1,0){5}}
\put(0,3.5){\line(0,1){3}}
\put(1.5,6){$\A$}
\multiput(5,2.5)(40,0){1}{\framebox(5,5){$\U$}}
\put(10,3){\line(1,0){8}}
\put(10,7){\line(1,0){8}}
\put(11,7.5){$\sb \in \SSS$}
\put(11,3.5){$\sb' \in \SSS$}
\multiput(18,6)(40,0){1}{\framebox(2,2){$+$}}
\put(19,6){\line(0,-1){2}}
\put(20,7){\line(1,0){7}}
\put(20.5,4.5){$\tb \in \SSS$}
\multiput(18,2)(40,0){1}{\framebox(2,2){$+$}}
\put(20,3){\line(1,0){7}}
\put(27,6.3){$\square$}
\put(27,2.3){$\square$}
\put(6,0){(e)}

\put(40,45){\line(1,0){5}}
\put(40,43.5){\line(0,1){3}}
\put(41.5,46){$\hat{\A}$}
\multiput(45,42.5)(40,0){1}{\framebox(5,5){$\U^\perp$}}
\put(50,43){\line(1,0){9}}
\put(50,47){\line(1,0){9}}
\put(51,47.5){$\hat{\sb} \in \hat{\SSS}$}
\put(58.5,46){$\circ$}
\put(51,43.5){$\hat{\sb}' \in \hat{\SSS}$}
\put(58.5,43){$\circ$}
\multiput(58,44)(40,0){1}{\framebox(2,2){$+$}}
\put(46,40){(f)}

\put(40,35){\line(1,0){5}}
\put(40,33.5){\line(0,1){3}}
\put(41.5,36){$\hat{\A}$}
\multiput(45,32.5)(40,0){1}{\framebox(5,5){$\U^\perp$}}
\put(50,33){\line(1,0){9}}
\put(50,37){\line(1,0){9}}
\put(51,37.5){$\hat{\sb} \in \hat{\SSS}$}
\put(58.5,36){$\circ$}
\put(51,33.5){$\hat{\sb}' \in \hat{\SSS}$}
\put(58.5,33){$\circ$}
\multiput(58,34)(40,0){1}{\framebox(2,2){$+$}}
\put(60,35){\line(1,0){7}}
\put(61,35.5){$\hat{\tb} \in \hat{\SSS}$}
\put(67,34.3){$\square$}
\put(46,30){(g)}

\put(40,25){\line(1,0){5}}
\put(40,23.5){\line(0,1){3}}
\put(41.5,26){$\hat{\A}$}
\multiput(45,22.5)(40,0){1}{\framebox(5,5){$\U^\perp$}}
\put(50,23){\line(1,0){9}}
\put(50,27){\line(1,0){9}}
\put(51,27.5){$\hat{\sb} \in \hat{\SSS}$}
\put(59,27){\line(0,-1){1}}
\put(51,23.5){$\hat{\sb}' \in \hat{\SSS}$}
\put(58.5,23){$\circ$}
\multiput(58,24)(40,0){1}{\framebox(2,2){$=$}}
\put(60,25){\line(1,0){7}}
\put(61,25.5){$\hat{\tb} \in \hat{\SSS}$}
\put(67,24.5){$\blacksquare$}
\put(46,20){(h)}

\put(40,15){\line(1,0){5}}
\put(40,13.5){\line(0,1){3}}
\put(41.5,16){$\hat{\A}$}
\multiput(45,12.5)(40,0){1}{\framebox(5,5){$\U^\perp$}}
\put(50,13){\line(1,0){8}}
\put(50,17){\line(1,0){8}}
\put(51,17.5){$\hat{\sb} \in \hat{\SSS}$}
\put(51,13.5){$\hat{\sb}' \in \hat{\SSS}$}
\put(60.5,14.5){$\hat{\tb} \in \hat{\SSS}$}
\multiput(58,16)(40,0){1}{\framebox(2,2){$+$}}
\put(59,16){\line(0,-1){1}}
\put(60,17){\line(1,0){7}}
\put(58.5,14){$\circ$}
\multiput(58,12)(40,0){1}{\framebox(2,2){$+$}}
\put(60,13){\line(1,0){7}}
\put(67,16.3){$\square$}
\put(67,12.3){$\square$}
\put(46,10){(i)}

\put(40,5){\line(1,0){5}}
\put(40,3.5){\line(0,1){3}}
\put(41.5,6){$\hat{\A}$}
\multiput(45,2.5)(40,0){1}{\framebox(5,5){$\U^\perp$}}
\put(50,3){\line(1,0){8}}
\put(50,7){\line(1,0){8}}
\put(51,7.5){$\hat{\sb} \in \hat{\SSS}$}
\put(51,3.5){$\hat{\sb}' \in \hat{\SSS}$}
\multiput(58,6)(40,0){1}{\framebox(2,2){$=$}}
\put(59,6){\line(0,-1){1}}
\put(60,7){\line(1,0){7}}
\put(58.5,4){$\circ$}
\multiput(58,2)(40,0){1}{\framebox(2,2){$=$}}
\put(60,3){\line(1,0){7}}
\put(60.5,4.5){$\hat{\tb} \in \hat{\SSS}$}
\put(67,6.5){$\blacksquare$}
\put(67,2.5){$\blacksquare$}
\put(46,0){(j)}

\end{picture}
\caption{Five pairs of dual normal realizations of $\CC$ and $\CC^\perp$.}
\label{NRD}
\end{figure}
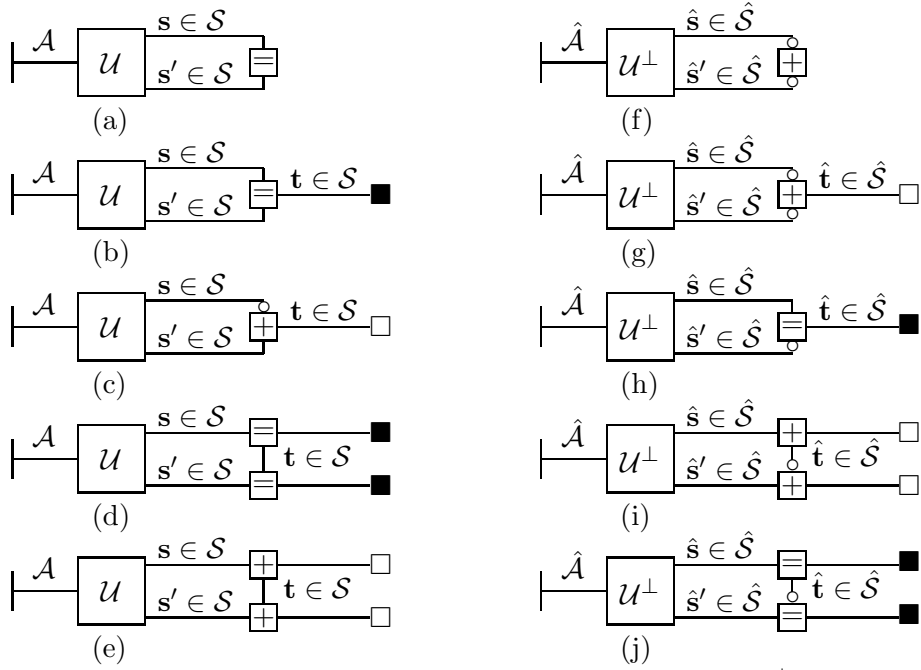

To illustrate this theorem, we show five pairs of dual normal realizations  in Figure \ref{NRD}.  Figure \ref{NRD}(a) realizes $\CC$ as the set of all $\ab \in \A$ such that there exists some $(\ab, \sb, \sb) \in \U$.  Figure \ref{NRD}(b) realizes $\CC$ as the projection $\Bf_{|\A}$ of the behavior $\Bf$.    Figure \ref{NRD}(c) realizes $\CC$ as the set of all $\ab \in \A$ such that there exists some $(\ab, \sb, \sb') \in \U$ whose \emph{syndrome} $\tb = \sb - \sb'$ is zero.    Figure \ref{NRD}(d) realizes $\CC$ as the projection $\bar{\Bf}_{|\A}$ of the extended behavior $\bar{\Bf}$.    Figure \ref{NRD}(e) realizes $\CC$ as the cross-section $\{(\ab, \sb + \tb, \sb' + \tb) : (\ab, \sb, \sb') \in \U, \tb \in \SSS\}_{:\A}$;  or, more succinctly, as $(\U + \W)_{:\A}$, where $\W = \{(\zerob, \sb, \sb) : \sb \in \SSS\}$.  Any of these realizations is evidently equivalent to Figure \ref{NRD}(a), so all are equivalent.

By dualizing Figures \ref{NRD}(a)--(e), we obtain the dual realizations shown in Figures \ref{NRD}(f)--(j) (where certain inessential sign inverters have been omitted).\footnote{
Realizations involving sign-inversion constraints may sometimes be simplified by the following rules: (a) since $\CC_i = -\CC_i$ if $\CC_i$ is abelian, a constraint $\CC_i$ is unchanged if sign-inversion constraints are added to all incident half-edges; (b) the cascade of two sign-inversion constraints is equivalent to an equality constraint, or simply to an edge.}
Figure \ref{NRD}(f) realizes $\CC^\circ$ as the set of all $\hat{\ab} \in \hat{\A}$ such that there exists some $(\hat{\ab}, \hat{\sb}, -\hat{\sb}) \in \U^\perp$.    Figure \ref{NRD}(g) realizes $\CC^\circ$ as the set of all $\hat{\ab} \in \hat{\A}$ such that there exists some $(\hat{\ab}, \hat{\sb}, \hat{\sb}') \in \U^\perp$ whose syndrome $\hat{\tb} = \hat{\sb} + \hat{\sb}'$ is zero.   Figure \ref{NRD}(h) realizes $\CC^\circ$ as the projection $(\Bf^\circ)_{|\hat{\A}}$ of the dual behavior $\Bf^\circ$.    Figure \ref{NRD}(i) realizes $\CC^\circ$ as the cross-section $(\U^\perp + \V^\perp)_{:\A}$, where $\V^\perp = \{(\zerob, \hat{\sb}, -\hat{\sb}) : \hat{\sb} \in \hat{\SSS}\}$ is the validity check space defined above.  Figure \ref{NRD}(j) realizes $\CC^\circ$ as the projection $(\bar{\Bf}^\circ)_{|\hat{\A}}$ of the extended dual behavior $\bar{\Bf}^\circ$.  Any of these realizations is evidently equivalent to Figure \ref{NRD}(f).

Our proof of the normal realization duality theorem follows from  Figures \ref{NRD}(d) and (i), which show that $\CC = \bar{\Bf}_{|\A} = (\U \cap \V)_{|\A}$  and $\CC^\circ = (\U^\perp + \V^\perp)_{:\hat{\A}}$, respectively, so $\CC^\perp = \CC^\circ$ by projection/cross-section and sum/intersection duality.  We could have equally well used Figures \ref{NRD}(e) and (j), which show that $\CC = (\U + \W)_{:\A}$  and $\CC^\circ = (\U^\perp \cap \W^\perp)_{|\hat{\A}}$, respectively.

  Figures \ref{NRD}(b), (d), (h) and (j) realize $\CC$ and $\CC^\perp$ as projections, sometimes called \emph{image} or \emph{generator representations}, whereas Figures \ref{NRD}(c), (e), (g) and (i) realize $\CC$ and $\CC^\perp$ as cross-sections, sometimes called \emph{kernel realizations}.  In particular, we note from Figure \ref{NRD}(c) that the extended behavior $\bar{\Bf}$ is the kernel of the \emph{syndrome-former homomorphism} $\U \to \SSS$ defined by $\tb = \sb - \sb'$.

Finally, we may straightforwardly generalize normal realization duality  to fragments as follows.
Since a fragment $\FF$ is a normal realization of its external behavior $\CC^\FF$, we  define   its \emph{dual fragment}  $\FF^\circ$ to be the dual normal realization to $\FF$.   Then, by normal realization duality, $\FF^\circ$ realizes $(\CC^\FF)^\perp$.

\subsection{Observability and controllability of realizations}\label{OCR}

We now discuss the  properties of observability and controllability of realizations as defined in \cite{FGL12}.  Using the ideas of the previous section, we obtain an elegant proof of observability/controllability duality, and a nice generalization of the controllability test of \cite{FGL12}.

A linear or group realization $\RR$ is said to be \emph{observable}, or \emph{one-to-one},  if the projection $\Bf \to \CC$ is one-to-one;  \ie  if the symbol configuration $\ab \in \CC$ determines the  internal state configuration $\sb \in \SSS$.     Evidently $\RR$ is  observable if and only if the \emph{unobservable state configuration space} $\SSS^{u} =  \Bf_{:\SSS}$ is trivial, since $\SSS^{u}$ is isomorphic to the kernel $\Bf^u = \{(\zerob, \sb) \in \Bf\}$ of this projection.  Alternatively, $\RR$ is  observable if and only if $|\Bf| = |\CC|$, since $\CC = \Bf_{|\A}$;  otherwise $|\Bf| > |\CC|$.

Equivalently, $\RR$ is observable if the projection $\bar{\Bf} \to \CC$ is one-to-one.  The kernel of this projection, $\bar{\Bf}^u = \{(\zerob, \sb, \sb) \in \bar{\Bf}\}$, is evidently isomorphic to $\Bf^u \cong \SSS^{u}$.

Figure \ref{OCD}(a) shows a normal realization of the unobservable state configuration space $\SSS^u$ as the cross-section $\Bf_{:\SSS}$.  Figure \ref{OCD}(d) shows a normal realization of the dual unobservable state configuration space $\hat{\SSS}^u$ as the cross-section $(\Bf^\circ)_{:\hat{\SSS}}$.

\begin{figure}[h]
\setlength{\unitlength}{5pt}
\centering
\begin{picture}(64,16)(-2, 2)
\put(0,15){\line(1,0){5}}
\put(-1.5,14.3){$\square$}
\put(1.5,16){$\A$}
\multiput(5,12.5)(40,0){1}{\framebox(5,5){$\U$}}
\put(10,13){\line(1,0){9}}
\put(10,17){\line(1,0){9}}
\put(11,17.5){$\sb \in \SSS$}
\put(19,17){\line(0,-1){1}}
\put(11,13.5){$\sb' \in \SSS$}
\put(19,13){\line(0,1){1}}
\multiput(18,14)(40,0){1}{\framebox(2,2){$=$}}
\put(20,15){\line(1,0){7}}
\put(21,15.5){$\tb \in \SSS$}
\put(6,10){(a)}

\put(33.5,14.3){$\blacksquare$}
\put(35,15){\line(1,0){5}}
\put(36.5,16){$\hat{\A}$}
\multiput(40,12.5)(40,0){1}{\framebox(5,5){$\U^\perp$}}
\put(45,13){\line(1,0){9}}
\put(45,17){\line(1,0){9}}
\put(46,17.5){$\hat{\sb} \in \hat{\SSS}$}
\put(46,13.5){$\hat{\sb}' \in \hat{\SSS}$}
\put(53.5,16){$\circ$}
\multiput(53,14)(40,0){1}{\framebox(2,2){$+$}}
\put(53.5,13.0){$\circ$}
\put(55,15){\line(1,0){7}}
\put(56,15.5){$\hat{\tb} \in \hat{\SSS}$}
\put(41,10){(c)}

\put(-1.5,4.3){$\blacksquare$}
\put(0,5){\line(1,0){5}}
\put(1.5,6){$\A$}
\multiput(5,2.5)(40,0){1}{\framebox(5,5){$\U$}}
\put(10,3){\line(1,0){9}}
\put(10,7){\line(1,0){9}}
\put(11,7.5){$\sb \in \SSS$}
\put(54,7){\line(0,-1){1}}
\put(11,3.5){$\sb' \in \SSS$}
\put(19,3){\line(0,1){1}}
\multiput(18,4)(40,0){1}{\framebox(2,2){$+$}}
\put(20,5){\line(1,0){7}}
\put(21,5.5){$\tb \in \SSS$}
\put(6,0){(b)}

\put(33.5,4.3){$\square$}
\put(35,5){\line(1,0){5}}
\put(36.5,6){$\hat{\A}$}
\multiput(40,2.5)(40,0){1}{\framebox(5,5){$\U^\perp$}}
\put(45,3){\line(1,0){9}}
\put(45,7){\line(1,0){9}}
\put(46,7.5){$\hat{\sb} \in \hat{\SSS}$}
\put(46,3.5){$\hat{\sb}' \in \hat{\SSS}$}
\put(18.5,6){$\circ$}
\multiput(53,4)(40,0){1}{\framebox(2,2){$=$}}
\put(53.5,3.0){$\circ$}
\put(55,5){\line(1,0){7}}
\put(56,5.5){$\hat{\tb} \in \hat{\SSS}$}
\put(41,0){(d)}
\end{picture}
\caption{Dual normal realizations of (a) $\SSS^u$; (b) $\SSS^c$; (c) $\hat{\SSS}^c$;  (d) $\hat{\SSS}^u$.}
\label{OCD}
\end{figure}

 In general, for any notion of observability that we encounter in this paper, we will define a dual property, which we will call ``controllability,"  such that  a linear or group fragment $\FF$ is controllable if and only if the dual fragment $\FF^\circ$ is observable.  As we will see, this property may or may not  correspond to classical notions of controllability in linear systems theory.

In the current context, we will say that a realization $\RR$ is \emph{controllable} if the two check subspaces $\U^\perp = \prod_{i \in \I_\CC} (\CC_i)^\perp$ and $\V^\perp = \{(\zerob, \hat{\sb}, -\hat{\sb}) \in \hat{\A} \times \hat{\SSS} \times \hat{\SSS}\}$ are independent;  \ie if $\U^\perp \cap \V^\perp = \{0\}$, so the check space $\bar{\Bf}^\perp = \U^\perp + \V^\perp$ is  an internal direct product, $\bar{\Bf}^\perp = \U^\perp \times \V^\perp$.  

\pagebreak
Now
$\U^\perp \cap \V^\perp = \{(\zerob, \hat{\sb}, -\hat{\sb}) \in \U^\perp\}$, 
 the unobservable extended behavior $(\bar{\Bf}^\circ)^u$ of the dual realization $\RR^\circ$.  Thus $\RR$ is controllable if and only if $\RR^\circ$ is observable.  This gives a succinct proof of the observability/controllability duality theorem for linear or group realizations  \cite[Theorem 4]{FGL12}. 

We note that by sum/intersection duality $\U^\perp \cap \V^\perp = \{0\}$ if and only if $\U + \V = \A \times \SSS \times \SSS$;  \ie $\RR$ is controllable if and only if $\U + \V = \A \times \SSS \times \SSS$.

The dual normal realization to that of Figure \ref{OCD}(a) is that of Figure \ref{OCD}(c), 
which realizes what we will call the \emph{controllable subspace} $\hat{\SSS}^c \subseteq \hat{\SSS}$ of the dual realization $\RR^\circ$, namely the set of syndromes $\hat{\tb} = \hat{\sb} + \hat{\sb}' \in \hat{\SSS}$ that occur as $(\hat{\ab}, \hat{\sb}, \hat{\sb}')$ runs through $\U^\perp$.  Similarly, the dual normal realization to that of Figure \ref{OCD}(d) is that of Figure \ref{OCD}(b), 
which realizes  the controllable subspace $\SSS^c \subseteq \SSS$ of the primal realization $\RR$, namely the set of syndromes $\tb = \sb - \sb' \in \SSS$ that occur as $(\ab, \sb, \sb')$ runs through $\U$.  By normal realization duality, we have immediately:

\vspace{1ex}
\noindent
\textbf{Theorem} (\textbf{unobservable state configuration space/controllable subspace duality}). The  unobservable state configuration  space $\SSS^u$ of a linear or group normal realization $\RR$ and the controllable subspace $\hat{\SSS}^c$ of its dual $\RR^\circ$ are orthogonal;  \ie  $\hat{\SSS}^c = (\SSS^u)^\perp$.  Similarly, $\SSS^c = (\hat{\SSS}^u)^\perp$. Thus $\RR$ (resp.\ $\RR^\circ$) is controllable if and only if $\SSS^c = \SSS$ (resp.\ $\hat{\SSS}^c = \hat{\SSS}$).  \qed \vspace{1ex}

We may take $|\SSS^u|$ or $\dim \SSS^u$ as a measure of the unobservability of $\RR$.  It follows from this result that $\SSS^u$ acts as the dual group or space to $\hat{\SSS}/\hat{\SSS}^c$, which in our setting implies that $\SSS^u \cong \hat{\SSS}/\hat{\SSS}^c$.  Thus $|\SSS^u| = |\hat{\SSS}|/|\hat{\SSS}^c|$;  or, in the linear case, $\dim \SSS^u = \dim \hat{\SSS} - \dim \hat{\SSS}^c$.  Thus if we take $|\hat{\SSS}|/|\hat{\SSS}^c|$ or $\dim \hat{\SSS} - \dim \hat{\SSS}^c$ as a measure of the uncontrollability of $\RR^\circ$, then  this theorem says that these measures of the unobservability of $\RR$ and the uncontrollability of $\RR^\circ$  are  ``the same size."

As we have seen, $\SSS^c$ is the image of the syndrome-former homomorphism $\U \to \SSS$ defined by $(\ab, \sb, \sb') \mapsto \sb - \sb'$, whose kernel is the extended behavior $\bar{\Bf}$.  By the fundamental theorem of homomorphisms, we  have $\U/\bar{\Bf} \cong \SSS^c$.  We 
therefore obtain the following generalization of the controllability test of \cite[Theorem 6]{FGL12}:

\vspace{1ex}
\noindent
\textbf{Theorem} (\textbf{controllability test}).  For a linear or group realization $\RR$ with extended behavior $\bar{\Bf} \subseteq \U$ and controllable subspace $\SSS^c \subseteq \SSS$, we have $|\U|/|\bar{\Bf}| = |\SSS^c| \le |\SSS|$, or in the linear case $\dim \U - \dim \bar{\Bf} = \dim \SSS^c \le \dim \SSS$,
with equality if and only if $\RR$ is controllable.  \qed \vspace{1ex}

In other words, a realization is uncontrollable if and only if its internal behavior is redundant in the following sense:
 $$|\bar{\Bf}| > \frac{|\U|}{|\SSS|} = \frac{\prod_i |\CC_i|}{\prod_j |\SSS_j|},$$
 or, in the linear case,
 $\dim \bar{\Bf} > \dim \U - \dim \SSS = \sum_i \dim \CC_i - \sum_j \dim \SSS_j.$
 
 As discussed in \cite{FGL12}, it would seem that it would always be desirable for iterative decoding to use observable (one-to-one) realizations.  However, is controllability always advantageous?  It is easy to see that a parity-check realization (\eg an LDPC code realization) is always observable, and is controllable if and only if its parity checks are independent \cite{FGL12}.  But redundant parity checks have some theoretical advantages, and have sometimes been used in practice.  Thus a judicious use of a bit of uncontrollability may sometimes be helpful.\footnote{At the oral presentation of \cite{F12}, John Baras made the interesting comment that in control systems design, a little bit of uncontrollability is sometimes used for robustness, even at the cost of some nonminimality.}
 
Finally, since a fragment $\FF$ may be regarded as a normal realization of its external behavior $\CC^\FF$, 
this development applies also to fragments.  For fragments, we will refer to this kind of observability and controllability as \emph{internal observability} and \emph{internal controllability}  (see Sections \ref{OF} and \ref{CF}).

\subsection{Generalized normal realization duality}\label{GEC}

The normal realization duality theorem may be generalized so as to exhibit greater symmetry between  primal and dual realizations as follows. 

The  primal generalized realization $\RR$ is defined much as before, but with extended behavior $\bar{\Bf} = \{(\ab, \sb, \sb') \in \U : s_j' = \varphi_j(s_j), \forall j \in \I_\SSS\}$, where, for each $j$, $\varphi_j:  \SSS_j \to \SSS_j'$ is an isomorphism between state spaces $\SSS_j$ and $\SSS_j'$.  The dual generalized realization $\RR^\circ$ is then defined with extended behavior $\bar{\Bf}^\circ = \{(\hat{\ab}, \hat{\sb}, \hat{\sb}') \in \U^\perp : \hat{s}_j = -\hat{\varphi}_j(\hat{s}_j'), \forall j \in \I_\SSS\}$, where $\hat{\varphi}_j:  \hat{\SSS}_j' \to \hat{\SSS}_j$ is the \emph{adjoint isomorphism} to $\varphi_j$ (see Section \ref{AIS}).
The constraint codes $C_j = \{(s_j, \varphi_j(s_j)) : s_j \in \SSS_j\}$ and $(C_j)^\perp = \{(-\hat{\varphi}_j(\hat{s}_j'), \hat{s}_j') : \hat{s}_j' \in \hat{\SSS}_j'\}$ are then  orthogonal for each $j \in \I_\SSS$, so by normal realization duality the codes realized by $\RR$ and $\RR^\circ$ are orthogonal.

\begin{figure}[h]
\setlength{\unitlength}{5pt}
\centering
\begin{picture}(55,6)(-2, 2)
\put(0,5){\line(1,0){5}}
\put(0,3.5){\line(0,1){3}}
\put(1.5,6){$\A$}
\multiput(5,2.5)(40,0){1}{\framebox(5,5){$\U$}}
\put(10,3){\line(1,0){9}}
\put(10,7){\line(1,0){9}}
\put(11,7.5){$\sb \in \SSS$}
\put(19,7){\line(0,-1){1}}
\put(11,3.5){$\sb' \in \SSS'$}
\put(19,3){\line(0,1){1}}
\multiput(17.5,4)(40,0){1}{\framebox(3,2){$\lra$}}
\put(6,0){(a)}
\put(30,5){\line(1,0){5}}
\put(30,3.5){\line(0,1){3}}
\put(31.5,6){$\hat{\A}$}
\multiput(35,2.5)(40,0){1}{\framebox(5,5){$\U^\perp$}}
\put(40,3){\line(1,0){9}}
\put(40,7){\line(1,0){9}}
\put(41,7.5){$\hat{\sb} \in \hat{\SSS}$}
\put(48.6,6.1){$\circ$}
\put(41,3.5){$\hat{\sb}' \in \hat{\SSS}'$}
\put(49,3){\line(0,1){1}}
\multiput(47.5,4)(40,0){1}{\framebox(3,2){$\hat{\lra}$}}
\put(36,0){(b)}
\end{picture}
\caption{Dual generalized normal realizations of $\CC$ and $\CC^\perp$.}
\label{GNRD}
\end{figure}
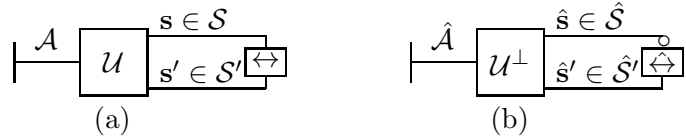

Figure \ref{GNRD} illustrates dual generalized  normal realizations of $\CC$ and $\CC^\perp$.  The box labeled by $\lra$ in Figure \ref{GNRD}(a) represents the set $\{s_j' = \varphi_j(s_j)\}$ of isomorphism constraints, whereas the box labeled by $\hat{\lra}$ and a small circle in Figure \ref{GNRD}(b) represents the set $\{\hat{s}_j = -\hat{\varphi}_j(\hat{s}_j')\}$ of negative adjoint isomorphism constraints.  

Moreover, we may correspondingly generalize normal graphs so as to exhibit greater symmetry between  primal and dual graphs as follows.  The primal generalized normal graph is defined much as before, except that the ends of each \emph{generalized edge}  represent values $s_j$ and $s_j'$ of isomorphic state spaces $\SSS_j$ and $\SSS_j'$, subject to some isomorphism constraint $s_j' = \varphi_j(s_j)$.  In the dual generalized normal graph, each dual generalized edge represents the negative adjoint isomorphism constraint $\hat{s}_j = -\hat{\varphi}_j(\hat{s}_j')$ between the dual state spaces $\hat{\SSS}_j$ and $\hat{\SSS}_j'$.  With such generalized edges, the primal and dual graphs  then have the same graph topology.

\section{External properties of fragments}\label{Section 4}

In this section we  begin our analysis of realizations via fragments.  Our main tool will be a simple but fundamental structure theorem for subdirect products, namely subgroups of an external direct product $A \times B$, or length-2 group codes.  We define trimness and properness for fragments, and generalize various results of \cite{FGL12} from constraint codes to fragments.  We show how fragments may be made trim and proper with respect to effective symbol variables.  Finally, we characterize the external state space for any trim and proper leaf fragment.  

\subsection{Fundamental theorem of subdirect products}\label{GCT}

We will use repeatedly the following fundamental result, which establishes the structure of any degree-2 linear or group constraint.  For further discussion, see \cite[Section VIII-D]{F01}. 

\pagebreak
\vspace{1ex}
\noindent
\textbf{Fundamental  theorem of subdirect products (FTSP)}.  Given groups $A$ and $B$ and a subgroup $C \subseteq A \times B$, let $C_{|A}$ and $C_{|B}$ be the projections of $C$ on $A$ and $B$, respectively, and let $C_{:A}$ and $C_{:B}$ be the cross-sections of $C$ on $A$ and $B$, respectively.  Then 
$$\frac{C_{|A}}{C_{:A}} \cong \frac{C_{|B}}{C_{:B}} \cong \frac{C}{C_{:A} \times C_{:B}}.$$

\vspace{1ex}
\noindent
\textsc{Proof}:  Evidently $C_{:A} \times C_{:B} \subseteq C$.  Since the kernels of the projections of $C$ and $C_{:A} \times C_{:B}$ on $B$ are both equal to $C_{:A}$, and their images are equal to $C_{|B}$ and $C_{:B}$, respectively, by the correspondence theorem we have $C/(C_{:A} \times C_{:B}) \cong C_{|B}/C_{:B}$.  Similarly, $C/(C_{:A} \times C_{:B}) \cong$ $C_{|A}/C_{:A}$.  \qed \vspace{1ex}

\begin{figure}[h]
\setlength{\unitlength}{5pt}
\centering
\begin{picture}(70,6)(-1, 2)
\multiput(-13,5)(50,0){1}{\line(1,0){4}}
\put(-12,6){$A$}
\multiput(-9,3)(50,0){1}{\line(0,1){4}}
\put(-8.5,5){$\hookleftarrow$}
\put(-8.5,4){$\rightarrow$}
\multiput(-9,3)(50,0){1}{\line(4,1){4}}
\multiput(-9,7)(50,0){1}{\line(4,-1){4}}
\multiput(-5,4)(50,0){1}{\line(0,1){2}}
\multiput(-5,5)(50,0){1}{\line(1,0){10}}
\multiput(5,3.5)(40,0){1}{\framebox(3,3){$\leftrightarrow$}}
\put(-4,6){$C_{|A}/C_{:A}$}
\put(9,6){$C_{|B}/C_{:B}$}
\multiput(22,5)(50,0){1}{\line(1,0){4}}
\put(23,6){$B$}
\multiput(22,3)(50,0){1}{\line(0,1){4}}
\put(19,5){$\hookrightarrow$}
\put(19,4){$\leftarrow$}
\multiput(22,3)(50,0){1}{\line(-4,1){4}}
\multiput(22,7)(50,0){1}{\line(-4,-1){4}}
\multiput(18,4)(50,0){1}{\line(0,1){2}}
\multiput(8,5)(30,0){1}{\line(1,0){10}}
\put(5,1){(a)}
\multiput(32,5)(50,0){1}{\line(1,0){4}}
\put(33,6){$\hat{A}$}
\multiput(36,3)(50,0){1}{\line(0,1){4}}
\put(36.5,5){$\hookleftarrow$}
\put(36.5,4){$\rightarrow$}
\multiput(36,3)(50,0){1}{\line(4,1){4}}
\multiput(36,7)(50,0){1}{\line(4,-1){4}}
\multiput(40,4)(50,0){1}{\line(0,1){2}}
\multiput(40,5)(50,0){1}{\line(1,0){14.3}}
\put(54,4.5){$\circ$}
\multiput(55,3.5)(40,0){1}{\framebox(3,3){$\hat{\leftrightarrow}$}}
\put(40.5,6){$(C_{:A})^\perp/(C_{|A})^\perp$}
\put(58.5,6){$(C_{:B})^\perp/(C_{|B})^\perp$}
\multiput(77,5)(50,0){1}{\line(1,0){4}}
\put(78,6){$\hat{B}$}
\multiput(77,3)(50,0){1}{\line(0,1){4}}
\put(74,5){$\hookrightarrow$}
\put(74,4){$\leftarrow$}
\multiput(77,3)(50,0){1}{\line(-4,1){4}}
\multiput(77,7)(50,0){1}{\line(-4,-1){4}}
\multiput(73,4)(50,0){1}{\line(0,1){2}}
\multiput(58,5)(30,0){1}{\line(1,0){15}}
\put(55,1){(b)}
\end{picture}
\caption{Realizations of dual subdirect products.}
\label{GC}
\end{figure}
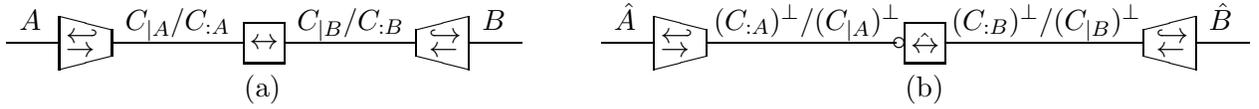

Figure \ref{GC}(a) shows a generic  realization of a subdirect product $C \subseteq A \times B$ according to this theorem.  
The first constraint  is $\{(a, a + C_{:A}) \in A \times C_{|A}/C_{:A} : a \in C_{|A}\}$.  This may be viewed as the combination of constraints based on the inclusion map $A \hookleftarrow C_{|A}$ and the natural map $C_{|A} \rightarrow C_{|A}/C_{:A}$, or equivalently constraints based on the natural map $A \to A/C_{:A}$ and the inclusion map $A/C_{:A} \hookleftarrow C_{|A}/C_{:A}$.  We will call such an inclusion/natural-map constraint on two variables an \emph{interface node}, as in \cite{F03}, and we will represent it by an isosceles trapezoid, which  indicates which of the two variable alphabets is smaller.

The central constraint in Figure \ref{GC}(a), depicted by a  box labeled by $\lra$,  is the isomorphism constraint $C_{|A}/C_{:A} \leftrightarrow C_{|B}/C_{:B}$.  (This represents   a generalized edge;  see Section \ref{GEC}.)  The final constraint is another interface node, namely $\{(b, b + C_{:B}) \in B \times C_{|B}/C_{:B} : b \in C_{|B}\}$.

Notice that  if we impose a zero constraint on $B$, then Figure \ref{GC}(a) realizes $C_{:A}$, whereas if we impose a dummy constraint on $B$, then Figure \ref{GC}(a) realizes $C_{|A}$.\footnote{More generally, with an appropriate constraint on $B$, Figure \ref{GC}(a) can realize the quotient  $D/C_{:A}$ for any $D$ such that $C_{:A} \subseteq D \subseteq C_{|A}$ is a normal series.  This is the gist of the ``most beautiful behavioral control theorem" \cite{T99, F01}.}
  
Figure \ref{GC}(b) shows the dual  realization of the orthogonal subdirect product $C^\perp \subseteq \hat{A} \times \hat{B}$. 
 Note that by projection/cross-section duality $(C^\perp)_{|\hat{A}} = (C_{:A})^\perp$ and $(C^\perp)_{:\hat{A}} = (C_{|A})^\perp$;  therefore, by quotient group duality, $(C^\perp)_{|\hat{A}}/(C^\perp)_{:\hat{A}}$ acts as the dual group to $C_{|A}/C_{:A}$.  Similarly, $(C^\perp)_{|\hat{B}}/(C^\perp)_{:\hat{B}}$ acts as the dual group to $C_{|B}/C_{:B}$.  
The dual isomorphism in Figure \ref{GC}(b) is thus the negative adjoint isomorphism to that  in Figure \ref{GC}(a);  see Section \ref{AIS}.

This dual realization and the FTSP imply that 
$$\frac{(C^\perp)_{|\hat{A}}}{(C^\perp)_{:\hat{A}}} \cong \frac{(C^\perp)_{|\hat{B}}}{(C^\perp)_{:\hat{B}}} \cong \frac{C^\perp}{(C^\perp)_{:\hat{A}} \times (C^\perp)_{:\hat{B}}}.$$

By quotient group duality, this implies
$$\frac{C_{|A}}{C_{:A}} \cong \frac{C_{|B}}{C_{:B}} \cong \frac{C_{|A} \times C_{|B}}{C},$$
which extends the FTSP to a ``fourth isomorphism"  in our setting.  
In summary, in the three normal series $C_{:A} \subseteq C_{|A}, C_{:B} \subseteq C_{|B}$ and $C_{:A} \times C_{:B} \subseteq C \subseteq  C_{|A} \times C_{|B}$, all four factor groups are isomorphic.

We caution that whereas the FTSP holds for general groups, since its proof depends only on the correspondence theorem, the ``fourth isomorphism" holds only for abelian groups.  Indeed, our proof holds only for finite abelian groups and vector spaces, where our duality theorems apply.

\begin{figure}[h]
\setlength{\unitlength}{5pt}
\centering
\begin{picture}(70,5)(-1, 2)
\multiput(-8,5)(50,0){1}{\line(1,0){4}}
\put(-7,6){$A$}
\multiput(-4,3)(50,0){1}{\line(0,1){4}}
\put(-3.5,4.5){$\rightarrow$}
\multiput(-4,3)(50,0){1}{\line(4,1){4}}
\multiput(-4,7)(50,0){1}{\line(4,-1){4}}
\multiput(0,4)(50,0){1}{\line(0,1){2}}
\multiput(0,5)(50,0){1}{\line(1,0){12}}
\multiput(12,3.5)(40,0){1}{\framebox(3,3){$\leftrightarrow$}}
\put(1,6){$A/(\ker \varphi)$}
\put(16,6){$\varphi(A)$}
\multiput(26,5)(50,0){1}{\line(1,0){4}}
\put(27,6){$B$}
\multiput(26,3)(50,0){1}{\line(0,1){4}}
\put(23,4.5){$\hookrightarrow$}
\multiput(26,3)(50,0){1}{\line(-4,1){4}}
\multiput(26,7)(50,0){1}{\line(-4,-1){4}}
\multiput(22,4)(50,0){1}{\line(0,1){2}}
\multiput(15,5)(30,0){1}{\line(1,0){7}}
\put(12,1){(a)}
\multiput(35,5)(50,0){1}{\line(1,0){4}}
\put(36,6){$\hat{A}$}
\multiput(39,3)(50,0){1}{\line(0,1){4}}
\put(39.5,4.5){$\hookleftarrow$}
\multiput(39,3)(50,0){1}{\line(4,1){4}}
\multiput(39,7)(50,0){1}{\line(4,-1){4}}
\multiput(43,4)(50,0){1}{\line(0,1){2}}
\multiput(43,5)(50,0){1}{\line(1,0){6.3}}
\put(49,4.5){$\circ$}
\multiput(50,3.5)(40,0){1}{\framebox(3,3){$\hat{\leftrightarrow}$}}
\put(44,6){$\hat{\varphi}(\hat{B})$}
\put(54,6){$\hat{B}/(\ker \hat{\varphi})$}
\multiput(69,5)(50,0){1}{\line(1,0){4}}
\put(70,6){$\hat{B}$}
\multiput(69,3)(50,0){1}{\line(0,1){4}}
\put(66,4.5){$\leftarrow$}
\multiput(69,3)(50,0){1}{\line(-4,1){4}}
\multiput(69,7)(50,0){1}{\line(-4,-1){4}}
\multiput(65,4)(50,0){1}{\line(0,1){2}}
\multiput(53,5)(30,0){1}{\line(1,0){12}}
\put(50,1){(b)}
\end{picture}
\caption{(a) Homomorphism $\varphi: A \to B$; (b) negative adjoint homomorphism $-\hat{\varphi}:  \hat{B} \to \hat{A}$.}
\label{GCH}
\end{figure}
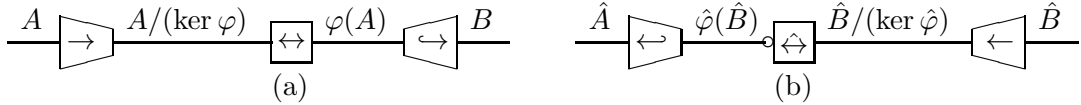

\vspace{1ex}
\noindent
\textbf{Remarks on homomorphisms}.
Given any homomorphism $\varphi:  A \to B$ with kernel $\ker \varphi$ and image $\varphi(A)$, the \emph{graph} of $\varphi$ is the subdirect product $C = \{(a, \varphi(a)) : a \in A\} \subseteq A \times B$.  Note that $C_{|A} = A, C_{:A} = \ker \varphi, C_{|B} = \varphi(A)$, and $C_{:B} = \{0\}$.  Thus we obtain the realization of $C$ shown in Figure \ref{GCH}(a),  where the first interface node is based on the natural map $A \rightarrow A/(\ker \varphi)$, and the last is based on the inclusion map $\varphi(A) \hookrightarrow B$.  The  fundamental theorem of homomorphisms, namely $\varphi(A) \cong A/(\ker \varphi)$, is thus a special case of the FTSP. 

\vspace{-1ex}
In this sense, a subdirect product $C \subseteq A \times B$ may be seen as a bidirectional generalization of a unidirectional homomorphism.  In systems theory terms, a homomorphism is an input-output (``cause-and-effect") system, whereas a subdirect product is a more general behavioral system. 

Moreover,  it is easily seen that the dual realization of Figure \ref{GCH}(b) represents the negative adjoint homomorphism $-\hat{\varphi}:  \hat{B} \to \hat{A}$ (see Section \ref{AIS}).
Thus an orthogonal subdirect product $C^\perp \subseteq \hat{A} \times \hat{B}$ generalizes a negative adjoint homomorphism.  \qed

\vspace{-1ex}

 \subsection{Trimness and properness for fragments}
 
 In \cite{FGL12} we defined trimness and properness for constraint codes,  showed that these were dual properties, and showed that lack of either of these properties at a state variable implies local reducibility.  In this section and the next we will straightforwardly generalize these results to fragments.

The external behavior $\CC^\FF$ of a fragment $\FF$ that involves an external state or symbol variable with alphabet $\V$ will be called \emph{trim} at $\V$ if the projection of $\CC^\FF$ on $\V$ is $\V$; \ie if the projection is surjective (onto).  $\FF$ will be called \emph{trim} if $\CC^\FF$ is trim at all its  variables.  Trimness is such an obviously desirable property that most authors assume  it, either implicitly or explicitly.

$\CC^\FF$ will be called \emph{proper} at $\V$ if the values of all other variables involved in $\CC^\FF$ determine the value $v \in \V$;  thus properness is a kind of local observability property.  If $\CC^\FF$ is a linear or group code, then it is proper at $\V$ if and only if zero values for all other variables imply $v = 0$;  \ie if and only if the cross-section $(\CC^\FF)_{:\V}$   is trivial.  $\FF$ will be called \emph{proper} if $\CC^\FF$ is proper at all its   variables;  \ie if none of the elements of $\CC^\FF$ has Hamming weight 1.
  
By projection/cross-section duality, $(\CC^\FF)_{|\V} = \V$ if and only if $((\CC^\FF)^\perp)_{:\hat{\V}} = \{0\}$.  Thus we have \emph{trim/proper duality}:  $\CC^\FF$ is trim at $\V$ if and only if the dual code $(\CC^\FF)^\perp$ is proper at $\hat{\V}$.  This is a straightforward generalization of  trim/proper duality for constraint codes \cite[Theorem 1]{FGL12}.  In view of this duality, we  conclude that trimness is a kind of local controllability property.

\vspace{1ex}
\noindent
\textbf{Remarks on homomorphisms} (cont.).  A subdirect product $C \subseteq A \times B$ is evidently the graph of a homomorphism if and only if it is trim at $A$ and proper at $B$.  The homomorphism is surjective (onto) if and only if it is  trim at $B$, and injective (one-to-one) if and only if it is  proper at $A$. $C$ is the graph of an isomorphism if and only if it is trim and proper at both $A$ and $B$.    \qed 

\subsection{Local reduction of state alphabets}\label{MTP}

We now show that if $\CC^\FF$ is not trim or proper at some variable $\V$, then the realization $\RR$ of which it is a part may be reduced if $\V$ is an external state variable, or effectively reduced if $\V$ is a symbol variable, with no essential change in graph topology. This generalizes  \cite[Theorem 2]{FGL12}.  

We partition the variables involved in $\FF$ into two subsets, one consisting of $\V$, and the other consisting of all other variables involved in $\FF$.  We denote the Cartesian product of all variable alphabets  involved in $\FF$ other than $\V$ by  $\bar{\V}^\FF$;  thus $\CC$ is a subgroup of the direct product $\V \times \bar{\V}^\FF$. By the  FTSP, $\CC^\FF$ then has the realization of Figure \ref{TPCp}.  

\begin{figure}[h]
\setlength{\unitlength}{5pt}
\centering
\begin{picture}(20,8)(-3, 2.5)
\multiput(-11,5)(50,0){1}{\line(1,0){7}}
\put(-9,6){$\bar{\V}^\FF$}
\multiput(-4,3)(50,0){1}{\line(0,1){4}}
\put(-3.5,5){$\hookleftarrow$}
\put(-3.5,4){$\rightarrow$}
\multiput(-4,3)(50,0){1}{\line(4,1){4}}
\multiput(-4,7)(50,0){1}{\line(4,-1){4}}
\multiput(0,4)(50,0){1}{\line(0,1){2}}
\multiput(0,5)(50,0){1}{\line(1,0){5}}
\multiput(5,3.5)(40,0){1}{\framebox(3,3){$\leftrightarrow$}}
\put(-5,2){\dashbox(14,7){}}
\put(7,10){$\tilde{\FF}$}
\put(1,6.5){$\tilde{\CC}^\FF$}
\multiput(17,5)(50,0){1}{\line(1,0){4}}
\put(18,6){$\V$}
\multiput(17,3)(50,0){1}{\line(0,1){4}}
\put(14,5){$\hookrightarrow$}
\put(14,4){$\leftarrow$}
\multiput(17,3)(50,0){1}{\line(-4,1){4}}
\multiput(17,7)(50,0){1}{\line(-4,-1){4}}
\multiput(13,4)(50,0){1}{\line(0,1){2}}
\multiput(8,5)(30,0){1}{\line(1,0){5}}
\put(10.5,6){$\tilde{\V}$}
\end{picture}
\caption{Realization of $\CC^\FF \subseteq \V \times \bar{\V}^\FF$.}
\label{TPCp}
\end{figure}

\vspace{-1ex}
Here we have introduced the \emph{reduced alphabet} $\tilde{\V} = (\CC^\FF)_{|\V}/(\CC^\FF)_{:\V}$.  In view of the normal series $\{0\} \subseteq (\CC^\FF)_{:\V} \subseteq (\CC^\FF)_{|\V} \subseteq \V$, we see that $|\tilde{\V}| \le |\V|$, with equality if and only if $\{0\} = (\CC^\FF)_{:\V}$ and $(\CC^\FF)_{|\V} = \V$;  \ie if and only if $\CC^\FF$ is trim and proper at $\V$.  

We have also introduced a \emph{reduced fragment} $\tilde{\FF}$ with \emph{effective external behavior} $\tilde{\CC}^\FF \subseteq \tilde{\V} \times \bar{\V}^\FF$.
We see that in any realization $\RR$ that includes $\FF$, we may replace $\FF$ by $\tilde{\FF}$ and the interface node between $\V$ and $\tilde{\V}$, namely $\{(v, v + (\CC^\FF)_{:\V}) \in \V \times \tilde{\V} : v \in (\CC^\FF)_{|\V}\}$.  

Moreover, if the variable $\V$ is an external state variable $\SSS_j$, we may then combine this interface node with the neighboring constraint code $\CC_{i}$ that also involves $\SSS_j$ to obtain an \emph{effective constraint code} $\tilde{\CC}_{i}$ that involves the \emph{reduced state variable} $\tilde{\SSS}_j = (\CC^\FF)_{|\SSS_j}/(\CC^\FF)_{:\SSS_j}$;  this amounts to restricting $\SSS_j$ to $(\CC^\FF)_{|\SSS_j}$ in $\CC_{i}$, and merging states $s_j \in (\CC^\FF)_{|\SSS_j}$ into their cosets $s_j + (\CC^\FF)_{:\SSS_j} \in \tilde{\SSS}_j$.  As a result, we obtain an equivalent realization $\tilde{\RR}$ with the same graph topology, but with $\FF, \SSS_j$ and $\CC_i$  reduced to $\tilde{\FF}, \tilde{\SSS}_j$ and $\tilde{\CC}_i$, as shown in Figure \ref{TPCr} (where the unlabeled edge represents the variables involved in $\CC_i$ other than $\SSS_j$).  As in \cite{FGL12}, we call this a \emph{local reduction} of $\RR$.

\begin{figure}[h]
\setlength{\unitlength}{5pt}
\centering
\begin{picture}(20,8)(3, 2.5)
\multiput(-11,5)(50,0){1}{\line(1,0){7}}
\put(-9,6){$\bar{\V}^\FF$}
\multiput(-4,3)(50,0){1}{\line(0,1){4}}
\put(-3.5,5){$\hookleftarrow$}
\put(-3.5,4){$\rightarrow$}
\multiput(-4,3)(50,0){1}{\line(4,1){4}}
\multiput(-4,7)(50,0){1}{\line(4,-1){4}}
\multiput(0,4)(50,0){1}{\line(0,1){2}}
\multiput(0,5)(50,0){1}{\line(1,0){5}}
\multiput(5,3.5)(40,0){1}{\framebox(3,3){$\leftrightarrow$}}
\put(-5,2){\dashbox(14,7){}}
\put(7,10){$\tilde{\FF}$}
\put(1,6.5){$\tilde{\CC}^\FF$}
\multiput(8,5)(50,0){1}{\line(1,0){5}}
\put(9.5,6){$\tilde{\SSS}_j$}
\multiput(17,3)(50,0){1}{\line(0,1){4}}
\put(14,5){$\hookrightarrow$}
\put(14,4){$\leftarrow$}
\multiput(17,3)(50,0){1}{\line(-4,1){4}}
\multiput(17,7)(50,0){1}{\line(-4,-1){4}}
\multiput(13,4)(50,0){1}{\line(0,1){2}}
\multiput(17,5)(50,0){1}{\line(1,0){5}}
\multiput(22,3)(40,0){1}{\framebox(4,4){$\CC_i$}}
\put(12,2){\dashbox(15,7){}}
\put(23,10){$\tilde{\CC}_i$}
\multiput(26,5)(30,0){1}{\line(1,0){5}}
\put(18.5,6){$\SSS_j$}
\end{picture}
\caption{Local reduction of $\FF, \SSS_j$ and $\CC_i$ to $\tilde{\FF}, \tilde{\SSS}_j$ and $\tilde{\CC}_i$.}
\label{TPCr}
\end{figure}

\vspace{-1ex}

  There are many definitions of minimality, but all have the property that a realization is not minimal if it has a local reduction of a single state space as above, with no change in graph topology or the size of other state spaces.  We therefore have, for any such definition of minimality: 

\vspace{1ex}
\noindent
\textbf{Theorem} (\textbf{minimal $\Rightarrow$ trim + proper}).
If a  linear or group normal realization $\RR$ is minimal, then  every constraint code $\CC_i$ is  trim and proper at all its  state variables. \qed \vspace{1ex}

Therefore, without loss of generality or minimality, we may and will assume that every constraint code $\CC_i$ is  trim and proper at all its  state variables.  If we are given a realization for which this assumption does not hold, then we may execute local reductions repeatedly until it does.  We shall see shortly that this simple iterative algorithm suffices to minimize any cycle-free realization.

We have already remarked that if a degree-2 constraint is trim and proper, then it is an isomorphism constraint.  If a degree-1 constraint is trim and proper, then it must be trivial. 

\vspace{1ex}
\noindent
\textbf{Remarks on sum-product decoding}.   The practical importance of a graphical representation of a code is that the graph may be used to specify a decoding algorithm.  The most common such algorithm is \emph{sum-product decoding} (also called ``belief propagation"), which is used to decode capacity-approaching codes such as low-density parity-check (LDPC) codes and turbo codes.  

The heart of the sum-product algorithm is as follows (see \eg \cite{F01, FV11, KFL01, L04, WLK95}).
      For each edge in a graph $\G$ representing a variable $\V$, the sum-product algorithm computes two ``messages" $\{\overrightarrow{\mu}(v), v \in \V\}$ and $\{\overleftarrow{\mu}(v), v \in \V\}$, corresponding to the two possible directions of the edge.  If the edge is incident on a vertex representing a constraint code $C$, then the ``outgoing" message is computed as a function of all of the ``incoming" messages on the other incident edges of $C$ by the  \emph{sum-product update rule}:
    $$
    \overrightarrow{\mu}(v) = \sum_{\cb \in C(v)} \prod_{\V' \neq \V} \overrightarrow{\mu}(\cb_{\V'}),
    $$
    where $C(v) = \{\cb \in C: \cb_\V = v\}$, the set of all $\cb \in C$ whose $\V$th component $\cb_\V$ is equal to $v$, and the product is over the other incoming message values $\overrightarrow{\mu}(\cb_{\V'})$ at the other  components $\cb_{\V'}$ of $\cb$. 

 If a constraint code $C$ is not trim at an incident variable $\V$, then the  message value $\overrightarrow{\mu}(v)$ computed by the sum-product update rule is evidently zero whenever $v \notin C_{|\V}$. Thus we may as well trim the message into a message over $C_{|\V}$, whether or not this local reduction has actually been performed. 
Dually, if a constraint code is not proper at $\V$, then it is easy to see that the message computed by the sum-product update rule satisfies   $\overrightarrow{\mu}(v) = \overrightarrow{\mu}(v + v')$  for any $v \in \V$, $v' \in C_{:\V}$;  \ie $\overrightarrow{\mu}(v)$ is constant over any coset of $C_{:\V}$.  Thus we may as well merge the message into a message over the cosets of $C_{:\V}$, whether or not this local reduction has actually been performed.  \qed

\subsection{Effective symbol alphabets}\label{ESA}

We now consider the case in which the variable $\V$ above is a symbol variable $\A_k$, and show  that  constraint codes may  be regarded as effectively trim and proper at symbol variables also. 

In this case, Figure \ref{TPCp}
becomes a realization of $\CC^\FF$ comprising an interface node between $\A_k$ and an  \emph{effective symbol variable} $\tilde{\A}_k = (\CC^\FF)_{|\A_k}/(\CC^\FF)_{:\A_k}$  (actually an internal state variable), and an effective constraint code $\tilde{\CC}^\FF$ involving $\tilde{\A}_k$ rather than $\A_k$.  The effective constraint code $\tilde{\CC}^\FF$ is trim and proper at $\tilde{\A}_k$, and $|\tilde{\A}_k| \le |\A_k|$, with equality if and only if $\CC^\FF$ is trim and proper at $\A_k$.

\begin{figure}[h]
\setlength{\unitlength}{5pt}
\centering
\begin{picture}(30,18)(-8, 4)
\multiput(-9,5)(50,0){1}{\line(1,0){7}}
\multiput(-9,3.5)(50,0){1}{\line(0,1){3}}
\put(-8,6){$\A_n$}
\multiput(-2,3)(50,0){1}{\line(0,1){4}}
\put(-1.5,5){$\hookleftarrow$}
\put(-1.5,4){$\rightarrow$}
\multiput(-2,3)(50,0){1}{\line(4,1){4}}
\multiput(-2,7)(50,0){1}{\line(4,-1){4}}
\multiput(2,4)(50,0){1}{\line(0,1){2}}
\multiput(2,5)(50,0){1}{\line(1,0){7}}
\put(3,6){$\tilde{\A}_n$}
\put(-1.5,9){$\cdots$}
\multiput(-9,15)(50,0){1}{\line(1,0){7}}
\multiput(-9,13.5)(50,0){1}{\line(0,1){3}}
\put(-8,16){$\A_2$}
\multiput(-2,13)(50,0){1}{\line(0,1){4}}
\put(-1.5,15){$\hookleftarrow$}
\put(-1.5,14){$\rightarrow$}
\multiput(-2,13)(50,0){1}{\line(4,1){4}}
\multiput(-2,17)(50,0){1}{\line(4,-1){4}}
\multiput(2,14)(50,0){1}{\line(0,1){2}}
\multiput(2,15)(50,0){1}{\line(1,0){7}}
\put(3,16){$\tilde{\A}_2$}
\multiput(-9,20)(50,0){1}{\line(1,0){7}}
\multiput(-9,18.5)(50,0){1}{\line(0,1){3}}
\put(-8,21){$\A_1$}
\multiput(-2,18)(50,0){1}{\line(0,1){4}}
\put(-1.5,20){$\hookleftarrow$}
\put(-1.5,19){$\rightarrow$}
\multiput(-2,18)(50,0){1}{\line(4,1){4}}
\multiput(-2,22)(50,0){1}{\line(4,-1){4}}
\multiput(2,19)(50,0){1}{\line(0,1){2}}
\multiput(2,20)(50,0){1}{\line(1,0){7}}
\multiput(9,2.5)(40,0){1}{\framebox(5,20){$\tilde{\CC}$}}
\put(3,21){$\tilde{\A}_1$}
\multiput(-4,3)(0,2){10}{\line(0,1){1}}
\multiput(7,3)(0,2){10}{\line(0,1){1}}
\end{picture}
\caption{Canonical decomposition of length-$n$ code $\CC \subseteq \prod_k \A_k$.}
\label{CD}
\end{figure}
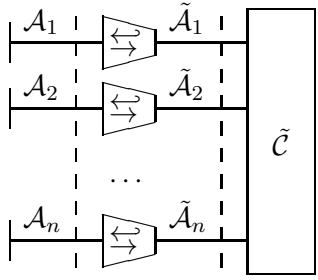

Given a linear or group realization $\RR$ of a length-$n$ code $\CC \subseteq  \prod_{k=1}^n \A_k$, this  decomposition may evidently be invoked for every symbol variable $\A_k$.  Thus we obtain a   decomposition of $\RR$ into  symbol variables, interface nodes, and a trim and proper constraint code $\tilde{\CC}$ whose realization $\tilde{\RR}$ has essentially the same graph topology as $\RR$, as illustrated in Figure \ref{CD}.  We note that the FTSP decomposition is a special case, in which $\tilde{\CC}$ is simply an isomorphism constraint.

\vspace{1ex}
\noindent
\textbf{Theorem} (\textbf{canonical decomposition}).  Any linear or group normal realization $\RR$ is equivalent to a realization $\tilde{\RR}$ with essentially the same graph topology consisting only of trim and proper constraint codes, plus interface nodes to symbol variables. \qed \vspace{1ex}

Consequently, we may   assume  that all constraints other than interface nodes involve only internal state variables, and furthermore are trim and proper at all  variables.

Moreover, we see that every external behavior $\CC^\FF$, including the  code $\CC$ realized by the entire realization $\RR$, may  be regarded as a ``coset code" over the effective symbol alphabets $\tilde{\A}_k$;  \ie over the cosets of the \emph{nondynamical symbol alphabet} $\underline{\A}_k = (\CC^\FF)_{:\A_k}$ in the \emph{trimmed symbol alphabet} $\bar{\A}_k = (\CC^\FF)_{|\A_k}$.  
Furthermore, if we choose to regard  the interface nodes as part of the external environment rather than of the realization, then we may consider any realization to be over its effective symbol alphabets $\tilde{\A}_k$, rather than over its symbol alphabets $\A_k$. 

We remark that if the nondynamical symbol alphabet $\underline{\A}_k$ is nontrivial, then the minimum distance between symbol configurations in $\CC$ cannot exceed the minimum distance within $\underline{\A}_k$, for any notion of distance, since every $a_k \in \underline{\A}_k$, combined with zeroes elsewhere, is a codeword in $\CC$.  
 
Dually, in any external behavior $\CC^\FF$, the symbol value $a_k$ must lie in the trimmed symbol alphabet $\bar{\A}_k$.  Because  symbol variable alphabets are fixed externally, we do not restrict  $\A_k$ to $\bar{\A}_k$, but rather let the interface node do the trimming.  Alternatively, if we regard  interface nodes as external, then the realization is effectively over the trimmed alphabet $\tilde{\A}_k$.

\vspace{1ex}
\noindent
\textbf{Remark on sum-product decoding}. Consider the sum-product update rule at an interface node between $\A_k$ and $\tilde{\A}_k$.  In the incoming message at $\tilde{\A}_k$, the weights of all $a_k$ in each coset of $\underline{\A}_k$ are simply combined to give the weight of that coset.  In the outgoing message at $\A_k$ (sometimes called the ``extrinsic information"), the weights of all symbols $a_k$  in each coset of $\underline{\A}_k$ are the same.  In other words, the extrinsic information  gives information  about $a_k$ only modulo $\underline{\A}_k$. \qed

\subsection{State space theorem for leaf fragments}\label{SAS}

In this section we consider a linear or group leaf fragment $\FF$;  \ie a fragment with only one external state variable $\SSS_j$.  The external behavior of such a fragment is then $\CC^\FF \subseteq \A^\FF \times \SSS_j$, where $\A^\FF = \prod_{A(\FF)} A_k$ is the symbol configuration space of $\FF$.  

By the FTSP, we  obtain the equivalent realization of $\FF$ shown in Figure \ref{SA}. 
 Here  $\tilde{\A}^\FF = \bar{\A}^\FF/\underline{\A}^\FF$ is the  \emph{effective symbol configuration space}, where $\bar{\A}^\FF = (\CC^\FF)_{|\A^\FF}$ is the \emph{trimmed symbol configuration space} and $\underline{\A}^\FF = (\CC^\FF)_{:\A^\FF}$ is the \emph{nondynamical symbol configuration space}.  The nondynamical  space $\underline{\A}^\FF$ thus comprises all symbol configurations $\ab^\FF \in \A^\FF$ that can occur with  $s_j = 0$, and the trimmed  space $\bar{\A}^\FF$ comprises all  $\ab^\FF \in \A^\FF$ that can occur with any $s_j \in \SSS_j$.  

\begin{figure}[h]
\setlength{\unitlength}{5pt}
\centering
\begin{picture}(30,3)(-9, 4)
\multiput(-13,5)(50,0){1}{\line(1,0){7}}
\multiput(-13,4)(50,0){1}{\line(0,1){2}}
\put(-11,6){$\A^\FF$}
\multiput(-6,3)(50,0){1}{\line(0,1){4}}
\put(-5.5,5){$\hookleftarrow$}
\put(-5.5,4){$\rightarrow$}
\multiput(-6,3)(50,0){1}{\line(4,1){4}}
\multiput(-6,7)(50,0){1}{\line(4,-1){4}}
\multiput(-2,4)(50,0){1}{\line(0,1){2}}
\multiput(-2,5)(50,0){1}{\line(1,0){7}}
\multiput(5,3.5)(40,0){1}{\framebox(3,3){$\leftrightarrow$}}
\put(0,6){$\tilde{\A}^\FF$}
\put(10,6){$\tilde{\SSS}_j$}
\multiput(19,5)(50,0){1}{\line(1,0){7}}
\put(21,6){$\SSS_j$}
\multiput(19,3)(50,0){1}{\line(0,1){4}}
\put(16,5){$\hookrightarrow$}
\put(16,4){$\leftarrow$}
\multiput(19,3)(50,0){1}{\line(-4,1){4}}
\multiput(19,7)(50,0){1}{\line(-4,-1){4}}
\multiput(15,4)(50,0){1}{\line(0,1){2}}
\multiput(8,5)(30,0){1}{\line(1,0){7}}
\end{picture}
\caption{Realization of leaf fragment external behavior $\CC^\FF \subseteq \A^\FF \times \SSS_j$.}
\label{SA}
\end{figure}

\begin{figure}[h]
\setlength{\unitlength}{5pt}
\centering
\begin{picture}(65,3)(4, 4)
\multiput(8,5)(14,0){1}{\line(1,0){8}}
\multiput(8,3.5)(30,0){1}{\line(0,1){3}}
\multiput(22,5)(14,0){1}{\line(1,0){6}}
\put(16,2.5){\framebox(6,5){$\CC^\FF$}}
\put(10,6){$\A^{\FF}$}
\put(25,6){$\SSS_j$}
\put(32,5){$=$}
\multiput(38,5)(14,0){1}{\line(1,0){7}}
\multiput(38,3.5)(30,0){1}{\line(0,1){3}}
\multiput(45,3)(30,0){1}{\line(0,1){4}}
\multiput(45,3)(30,0){1}{\line(4,1){4}}
\multiput(45,7)(30,0){1}{\line(4,-1){4}}
\multiput(49,4)(30,0){1}{\line(0,1){2}}
\put(45.5,5){$\hookleftarrow$}
\put(45.5,4){$\rightarrow$}
\put(40,6){$\A^{\FF}$}
\multiput(49,5)(14,0){1}{\line(1,0){7}}
\put(51,6){$\tilde{\A}^{\FF}$}
\put(56,3.5){\framebox(3,3){$\leftrightarrow$}}
\multiput(59,5)(14,0){1}{\line(1,0){6}}
\put(61,6){$\SSS_j$}
\end{picture}
\caption{State space theorem for a trim and proper leaf fragment $\FF$.}
\label{EqCp}
\end{figure}
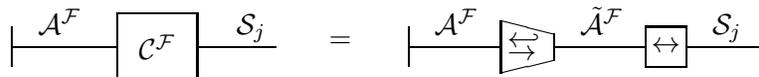

If $\CC^\FF$ is trim and proper at $\SSS_j$, then Figure \ref{SA}  reduces to Figure \ref{EqCp}, which illustrates the following simple but important theorem:

\pagebreak
\vspace{1ex}
\noindent
\textbf{Theorem}  (\textbf{state space theorem for leaf fragments}).  If the external behavior $\CC^\FF \subseteq \A^{\FF} \times \SSS_j$ of a linear or group leaf fragment $\FF$ is trim and proper at its external state space $\SSS_j$, then $\SSS_j$ is isomorphic to the effective symbol configuration space $\tilde{\A}^{\FF} = \bar{\A}^{\FF}/\underline{\A}^{\FF} = (\CC^\FF)_{|\A^{\FF}}/(\CC^\FF)_{:\A^{\FF}}$.

\vspace{1ex}
\noindent
\textsc{Proof}:  We have $\tilde{\A}^{\FF} \cong \tilde{\SSS}_j = (\CC^\FF)_{|\SSS_j}/(\CC^\FF)_{:\SSS_j}$ by the FTSP.  But if $\CC^\FF$ is trim and proper at $\SSS_j$, then $\tilde{\SSS_j} \cong \SSS_j$.  \qed 

\vspace{-2ex}
\section{Cycle-free and cyclic realizations}

In this section, we first  review some elementary graph theory.  Then, using the state space theorem for  leaf fragments, we  obtain an improved proof of the ``minimal $\Leftrightarrow$ trim and proper" theorem of \cite{FGL12}, which is the key result for cycle-free realizations.  Finally, we show that any trim and proper cyclic realization may be decomposed into a  leafless  ``2-core" and a number of cycle-free leaf fragments, to which this theorem  again  applies.

\subsection{Cycle-free and cyclic graphs}

Any finite graph $\G = (V,E)$ may be constructed by starting with the set $V$ of all its vertices, and then adding  the edges in its edge set $E$, one by one.  Thus initially there are $|V|$ disconnected components, each comprising one vertex and no edges.  Each added edge either connects two previously disconnected components, or creates a cycle  in some already connected component.

 The \emph{cyclomatic number}\footnote{Also called the circuit rank, cycle rank, nullity, or first Betti number.} of a graph $\G = (V,E)$ with $N_c$ connected components is defined as $N_\G = |E| - |V| + N_c$  \cite{Berge}.  Thus initially when $E$ is empty, we have $|E| = 0$ and $N_c = |V|$, so the cyclomatic number starts at zero.  If  adding an edge connects two disconnected components, then $|E|$ increases by 1 and $|N_c|$ decreases by 1, so $N_\G$ remains constant;  otherwise $N_\G$ increases by 1.  Thus $N_\G \ge 0$ for any graph $\G$, and $N_\G = 0$ if and only if $\G$ is cycle-free.  A connected graph $\G$ is thus cycle-free if and only if $|E| = |V| - 1$.  
 
 Moreover, a connected graph $\G$ is cycle-free if and only if every edge is a cut set;  that is, cutting any edge into two half-edges disconnects the graph into two components.  Each such component is a cycle-free leaf fragment, called a \emph{rooted tree} in graph theory, whose  \emph{root} is the associated half-edge.
 
 The \emph{degree} of a vertex is the number of incident edges.
 A  connected cycle-free graph has at least two leaf (degree-1) vertices.  A rooted tree has a least one leaf vertex, not counting  the root.
 
A graph $\G$ that is not cycle-free will be called \emph{cyclic}.   Its cyclomatic number $N_\G$ is then equal to the minimum number of edge cuts required to make $\G$ cycle-free, and also to the maximum number of edge cuts that can be made without disconnecting $\G$.  $N_\G$  thus  measures  the ``loopiness" of $\G$.
  
The \emph{2-core} of a connected  graph $\G$ is its maximal connected subgraph such that all vertices have degree 2 or more \cite{B84};  \ie the 2-core is the maximal connected leafless  subgraph. As in \cite{FGL12}, we will call a connected  leafless graph  a \emph{generalized cycle}. The 2-core of $\G$ may be found by repeatedly deleting leaf vertices until none remain.  The 2-core is empty if and only if $\G$ is cycle-free.
  
  Since the 2-core may be obtained from $\G$ by deleting leaf vertices and their associated edges, the cyclomatic number of the 2-core of $\G$ is the same as that of $\G$.  The 2-core of $\G$ thus comprises its essential cyclic skeleton after all leaves have been stripped away.

 \subsection{Connecting fragments}\label{CF1}
  
  We will now consider connecting a pair of disconnected fragments $\FF_1$ and $\FF_2$ to form a combined fragment $\FF_{12}$  by imposing an isomorphism constraint $\SSS_j \lra \SSS_j'$ on external state variables of $\FF_1$ and $\FF_2$, respectively.  In other words, we connect $\SSS_j$ and $\SSS'_j$ via a generalized edge.  When the isomorphism constraint is an equality constraint, this operation has been called ``closing the box"  \cite{VL03}.
Such a connection is illustrated in Figure \ref{linkCp}.  

\begin{figure}[h]
\setlength{\unitlength}{5pt}
\centering
\begin{picture}(30,6)(-5, 3)
\put(-17,5){\line(1,0){8}}
\multiput(-3,5)(50,0){1}{\line(1,0){8}}
\multiput(8,5)(30,0){1}{\line(1,0){8}}
\put(4.5,9.5){$\CC_{12}$}
\put(-10,1.5){\dashbox(33,7){}}
\multiput(-9,2.5)(40,0){1}{\framebox(6,5){$\CC_1$}}
\multiput(5,3.5)(40,0){1}{\framebox(3,3){$\lra$}}
\put(16,2.5){\framebox(6,5){$\CC_{2}$}}
\put(0,6){$\SSS_j$}
\put(11,6){$\SSS'_j$}
\put(22,5){\line(1,0){8}}
\end{picture}
\caption{Connecting two fragments via an isomorphism constraint $\SSS_j \lra \SSS'_j$.}
\label{linkCp}
\end{figure}

We then have the following simple but important lemma (to be continued in Section \ref{CFC}):

\vspace{1ex}
\noindent
\textbf{Lemma} (\textbf{connected fragments}).  If two linear or group fragments $\FF_1, \FF_2$  are  connected via a generalized edge between  state spaces $\SSS_j$ and $\SSS_j'$, and  $\FF_{12}$ is the combined fragment, then: \\
\indent (a) If $\FF_1$ and $\FF_2$ are  trim, then $\FF_{12}$ is  trim.  \\
\indent (b) If $\FF_1$ and $\FF_2$ are  proper, then $\FF_{12}$ is  proper.

\vspace{1ex}
\noindent
\textsc{Proof}:  (a) If $\FF_2$ is  trim, then every value $s_{j'}$ of every external state variable $\SSS_{j'}$ of $\FF_2$ appears in some valid configuration of $\CC_2$ in combination with some value $s'_j \in \SSS'_j$;  and if $\FF_1$ is  trim, then the corresponding value $s_j \in \SSS_j$ under the given isomorphism appears in some valid configuration of $\CC_1$, so  $s_{j'}$ must appear in some valid configuration in $\CC_{12}$;  and similarly for the  state variables of $\CC_1$.  Thus $\FF_{12}$ is trim.

(b) If all values of all variables of $\FF_{12}$ except any single state variable $\SSS_{j'}$ of $\FF_2$ are equal to zero, then $s_j = 0$ by the properness of $\FF_1$, so $s'_j = 0$ by the isomorphism, so $s_{j'} = 0$ by the properness of $\FF_2$;  and similarly for any state variable of $\CC_1$.  Thus $\FF_{12}$ is proper.  

Alternatively, (b)  follows from (a), or (a) from (b), by trim/proper duality. 
\qed \vspace{1ex}

We will  call a fragment $\FF$ \emph{internally trim} if all of its constraint codes are  trim, and \emph{internally proper} if all of its constraint codes are  proper.  These definitions generalize the corresponding definitions for realizations of \cite{FGL12}.

Any connected cycle-free graph may be constructed by starting with its vertices and iteratively connecting vertices via  edges as above.  Thus if all its constraint codes are trim (resp.\ proper), then by recursive application of the connected fragments lemma we have:

\vspace{1ex}
\noindent
\textbf{Theorem} (\textbf{trimness/properness of cycle-free fragments}).  If a cycle-free fragment is  internally trim (resp.\ proper), then it is  trim (resp.\ proper). \qed \vspace{1ex}

This theorem and the state space theorem for leaf fragments yield an important result:

\vspace{1ex}
\noindent
\textbf{Theorem}  (\textbf{cycle-free leaf fragments}).  If a linear or group cycle-free leaf fragment $\FF$ with external behavior $\CC^\FF \subseteq \A^\FF \times \SSS_j$  is internally trim and proper, then its external state space $\SSS_j$ is isomorphic to its effective symbol configuration space $\tilde{\A}^\FF = \bar{\A}^\FF/\underline{\A}^\FF$. \qed 

\subsection{Minimal cycle-free realizations}

We  now apply the cycle-free leaf fragment theorem to cycle-free realizations.  We  obtain an improved proof of one direction of the ``minimal = trim + proper" theorem of \cite[Theorem 3]{FGL12}. 

If $\RR$ is a cycle-free realization, then cutting any edge $\SSS_j$ into two half-edges disconnects $\RR$ into two cycle-free leaf fragments (rooted trees) $\FF_j$ and $\PP_j$, whose roots are these half-edges.  Then:

\vspace{1ex}
\noindent
\textbf{Lemma} (\textbf{trim + proper $\Rightarrow$ minimal}).
If a  finite connected linear or group normal realization $\RR$ is cycle-free and internally trim and proper, then every state space $\SSS_j$ is isomorphic to $\CC_{|\A^{\FF_j}}/\CC_{:\A^{\FF_j}}$, and also to $\CC_{|\A^{\PP_j}}/\CC_{:\A^{\PP_j}}$, where $\FF_j$ and $\PP_j$ are the two cycle-free leaf fragments of $\RR$ created by cutting the edge $\SSS_j$.  Moreover, $\SSS_j$ is minimal.

\begin{figure}[h]
\setlength{\unitlength}{5pt}
\centering
\begin{picture}(50,11)(36, 3)
\multiput(48,5)(14,0){1}{\line(1,0){7}}
\multiput(48,3.5)(30,0){1}{\line(0,1){3}}
\multiput(54,3)(0,2){6}{\line(0,1){1}}
\multiput(55,3)(30,0){1}{\line(0,1){4}}
\multiput(55,3)(30,0){1}{\line(4,1){4}}
\multiput(55,7)(30,0){1}{\line(4,-1){4}}
\multiput(59,4)(30,0){1}{\line(0,1){2}}
\put(55.5,5){$\hookleftarrow$}
\put(55.5,4){$\rightarrow$}
\put(49,6){$\A^{\PP_j}$}
\multiput(59,5)(14,0){1}{\line(1,0){7}}
\multiput(65,3)(0,2){6}{\line(0,1){1}}
\multiput(66,3.5)(40,0){1}{\framebox(3,3){$\leftrightarrow$}}
\put(60,6){$\tilde{\A}^{\PP_j}$}
\put(68,8){$\SSS_j$}
\multiput(67.5,6.5)(14,0){1}{\line(0,1){4}}
\multiput(66,10.5)(40,0){1}{\framebox(3,3){$\leftrightarrow$}}
\put(60,13){$\tilde{\A}^{\FF_j}$}
\multiput(59,12)(14,0){1}{\line(1,0){7}}
\multiput(48,12)(14,0){1}{\line(1,0){7}}
\multiput(48,10.5)(30,0){1}{\line(0,1){3}}
\multiput(55,10)(30,0){1}{\line(0,1){4}}
\multiput(55,10)(30,0){1}{\line(4,1){4}}
\multiput(55,14)(30,0){1}{\line(4,-1){4}}
\multiput(59,11)(30,0){1}{\line(0,1){2}}
\put(55.5,12){$\hookleftarrow$}
\put(55.5,11){$\rightarrow$}
\put(49,13){$\A^{\FF_j}$}
\end{picture}
\caption{Realization of $\CC$ as a subdirect product.}
\label{EQ2Cp}
\end{figure}
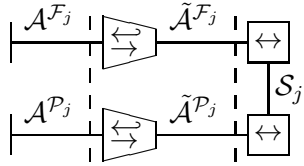

\vspace{1ex}
\noindent
\textsc{Proof}:  If $\RR$ is cycle-free, then both $\FF_j$ and $\PP_j$ are cycle-free leaf fragments with external state space $\SSS_j$, so by the cycle-free leaf fragment theorem, we have $\SSS_j \cong (\CC^{\FF_j})_{|\A^{\FF_j}}/(\CC^{\FF_j})_{:\A^{\FF_j}}$ and  $\SSS_j \cong (\CC^{\PP_j})_{|\A^{\PP_j}}/(\CC^{\PP_j})_{:\A^{\PP_j}}$.
Thus, as shown in Figure \ref{EQ2Cp}, $\CC$ must be the subdirect product 
$$\CC = \left\{(\ab^{\FF_j}, \ab^{\PP_j})  \subseteq \A^{\FF_j} \times \A^{\PP_j} :  \ab^{\FF_j} + (\CC^{\FF_j})_{:\A^{\FF_j}} \lra \ab^{\PP_j} + (\CC^{\PP_j})_{:\A^{\PP_j}} \right\},
$$
where $\lra$ denotes correspondence under the  isomorphisms
$$\tilde{\A}^{\FF_j} = \frac{(\CC^{\FF_j})_{|\A^{\FF_j}}}{(\CC^{\FF_j})_{:\A^{\FF_j}}} \cong \SSS_j \cong \frac{(\CC^{\PP_j})_{|\A^{\PP_j}}}{(\CC^{\PP_j})_{:\A^{\PP_j}}} = \tilde{\A}^{\PP_j}.$$
This implies $\CC_{|\A^{\FF_j}} = (\CC^{\FF_j})_{|\A^{\FF_j}}$, $\CC_{|\A^{\PP_j}} = (\CC^{\PP_j})_{|\A^{\PP_j}}$,  $\CC_{:\A^{\FF_j}} = (\CC^{\FF_j})_{:\A^{\FF_j}}$, and $\CC_{:\A^{\PP_j}} = (\CC^{\PP_j})_{:\A^{\PP_j}}$.

Moreover, in any realization with the same graph topology, the size of $\SSS_j$ must be at least $|\SSS_j|$, since if $\ab^{\FF_j}$ and $\ab^{\PP_j}$ are not in corresponding cosets of $(\CC^{\FF_j})_{:\A^{\FF_j}}$ and $(\CC^{\PP_j})_{:\A^{\PP_j}}$, then $(\ab^{\FF_j}, \ab^{\PP_j}) \notin \CC$.  Thus $\SSS_j$ is minimal.  
\qed \vspace{1ex}

We have already proved the converse (minimal $\Rightarrow$ trim + proper) in Section \ref{MTP}.  We have thus simplified  the proof of the following fundamental theorem:

\vspace{1ex}
\noindent
\textbf{Theorem} (\textbf{minimal $\Leftrightarrow$ trim + proper} \cite{FGL12}).
If a finite connected normal linear or group realization $\RR$ of a code $\CC$ is  cycle-free, then the following are equivalent:
\begin{itemize}
\item[(1)]  $\RR$ is internally trim and proper;
\item[(2)]  Every state space $\SSS_j$ is isomorphic to $\CC_{|\A^{\FF_j}}/\CC_{:\A^{\FF_j}}$, and also to $\CC_{|\A^{\PP_j}}/\CC_{:\A^{\PP_j}}$;
\item[(3)]  Every state space $\SSS_j$ is minimal; \ie $\RR$ is minimal. \qed
\end{itemize}

Parts (2) and (3) of this theorem are effectively the \textbf{state space theorem} of \cite{W89, FT93}.  

\subsection{Cycle-free leaf fragments and 2-cores}\label{TC}

We  now apply the cycle-free leaf fragment theorem to cyclic realizations.  

As we have seen in Section 6.1, a finite connected cyclic graph $\G$ has a unique maximal leafless subgraph $\bar{\G}$, called the 2-core of $\G$, which may be obtained from $\G$ by repeatedly deleting leaves, and which has the same cyclomatic number as $\G$.
In our context, we will define the \emph{2-core} $\bar{\RR}$ of a cyclic normal realization $\RR$ with normal graph $\G$ as the part of $\RR$ that remains after repeatedly deleting leaf constraints; thus the normal graph $\bar{\G}$ of $\bar{\RR}$ is the 2-core of $\G$.  

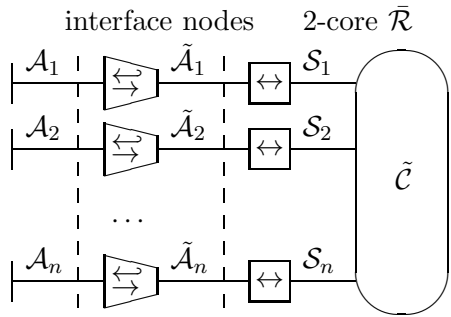
\begin{figure}[h]
\setlength{\unitlength}{5pt}
\centering
\begin{picture}(35,22)(-8, 4)
\multiput(-9,5)(50,0){1}{\line(1,0){7}}
\multiput(-9,3.5)(50,0){1}{\line(0,1){3}}
\put(-8,6){$\A_n$}
\multiput(-2,3)(50,0){1}{\line(0,1){4}}
\put(-1.5,5){$\hookleftarrow$}
\put(-1.5,4){$\rightarrow$}
\multiput(-2,3)(50,0){1}{\line(4,1){4}}
\multiput(-2,7)(50,0){1}{\line(4,-1){4}}
\multiput(2,4)(50,0){1}{\line(0,1){2}}
\multiput(2,5)(50,0){1}{\line(1,0){7}}
\put(3,6){$\tilde{\A}_n$}
\multiput(9,3.5)(40,0){1}{\framebox(3,3){$\leftrightarrow$}}
\multiput(12,5)(50,0){1}{\line(1,0){5}}
\put(13,6){$\SSS_n$}
\put(-1.5,9){$\cdots$}
\multiput(-9,15)(50,0){1}{\line(1,0){7}}
\multiput(-9,13.5)(50,0){1}{\line(0,1){3}}
\put(-8,16){$\A_2$}
\multiput(-2,13)(50,0){1}{\line(0,1){4}}
\put(-1.5,15){$\hookleftarrow$}
\put(-1.5,14){$\rightarrow$}
\multiput(-2,13)(50,0){1}{\line(4,1){4}}
\multiput(-2,17)(50,0){1}{\line(4,-1){4}}
\multiput(2,14)(50,0){1}{\line(0,1){2}}
\multiput(2,15)(50,0){1}{\line(1,0){7}}
\put(3,16){$\tilde{\A}_2$}
\multiput(9,13.5)(40,0){1}{\framebox(3,3){$\leftrightarrow$}}
\multiput(12,15)(50,0){1}{\line(1,0){5}}
\put(13,16){$\SSS_2$}
\multiput(-9,20)(50,0){1}{\line(1,0){7}}
\multiput(-9,18.5)(50,0){1}{\line(0,1){3}}
\put(-8,21){$\A_1$}
\multiput(-2,18)(50,0){1}{\line(0,1){4}}
\put(-1.5,20){$\hookleftarrow$}
\put(-1.5,19){$\rightarrow$}
\multiput(-2,18)(50,0){1}{\line(4,1){4}}
\multiput(-2,22)(50,0){1}{\line(4,-1){4}}
\multiput(2,19)(50,0){1}{\line(0,1){2}}
\multiput(2,20)(50,0){1}{\line(1,0){7}}
\multiput(20.5,12.5)(40,0){1}{\oval(7,20){}}
\put(20,12){$\tilde{\CC}$}
\put(3,21){$\tilde{\A}_1$}
\multiput(9,18.5)(40,0){1}{\framebox(3,3){$\leftrightarrow$}}
\multiput(12,20)(50,0){1}{\line(1,0){5}}
\put(13,21){$\SSS_1$}
\multiput(-4,3)(0,2){10}{\line(0,1){1}}
\multiput(7,3)(0,2){10}{\line(0,1){1}}
\put(-5,24){interface nodes}
\put(13,24){2-core $\bar{\RR}$}
\end{picture}
\caption{Schematic representation of 2-core $\bar{\RR}$ with $n$ cycle-free leaf fragments.}
\label{2CCq}
\end{figure}

The parts of $\RR$ that are stripped away then comprise a number of cycle-free leaf fragments (rooted trees),  shown schematically in Figure \ref{2CCq}.  Each such leaf fragment $\FF_i$ is connected to $\bar{\RR}$ via a single external state space (root) $\SSS_i$.  Under our standing assumption that all constraint codes are trim and proper, the cycle-free leaf fragment theorem applies to each such leaf fragment $\FF_i$, so $\SSS_i \cong \tilde{\A}_i = (\CC_i)_{|\A_i}/(\CC_i)_{:\A_i}$.  We  shall regard these isomorphism constraints (generalized edges) as parts of  $\bar{\RR}$.    (Note that the cycle-free case of Figure \ref{EQ2Cp} is a special case of Figure \ref{2CCq}.)

As in the canonical decomposition of Section \ref{ESA},
each cycle-free leaf fragment may be regarded as an interface node, and the 2-core $\bar{\RR}$ may be regarded as an effective realization of an effective code $\tilde{\CC}$ over the effective symbol configuration spaces $\tilde{\A}_i\cong \SSS_i$.
 
    Note that the boundary of the 2-core $\bar{\RR}$ consists entirely of effective symbol variables $\tilde{\A}_i$, and that all symbol variables $\A_i$ lie outside of this boundary.  
    Moreover, the effective code $\tilde{\CC}$ is trim and proper at each $\tilde{\A}_i$. The effective code $\tilde{\CC}$ may be lifted to $\CC$ by expanding the cosets of each $\underline{\A}_i$ to all of their elements. 

We thus have proved another useful decomposition theorem:

\vspace{1ex}
\noindent
\textbf{Theorem} (\textbf{second canonical  decomposition}).  An internally trim and proper,  cyclic, linear or group  realization $\RR$ may be decomposed into cycle-free leaf fragments with external state spaces $\tilde{\A}_i = \bar{\A}_i/\underline{\A}_i$, and a trim and proper  leafless 2-core $\bar{\RR}$ with effective symbol alphabets $\tilde{\A}_i$.  The cyclomatic number of $\bar{\RR}$ is the same as that of $\RR$. \qed \vspace{1ex}

\vspace{1ex}
\noindent
\textbf{Remarks}.
We expect that the main difficulties in code analysis and in decoding will be associated with the 2-core $\bar{\RR}$.
For example, in  sum-product decoding, decoding of  cycle-free leaf fragments is non-iterative and exact, with the result being a message of  weights of the elements of $\tilde{\A}_i$.  Iterative  sum-product decoding may then be performed on the 2-core graph $\bar{\G}$, with these incoming messages held constant.  As a simple example, in sum-product decoding of a tail-biting trellis code with parallel transitions, the  weights of the parallel transitions need to be computed only once. \qed

\section{Observability and controllability of fragments}\label{Section 5}

In this section, we will consider three kinds of observability and controllability for fragments, which we will call ``internal,"  ``external," and ``total."  Internal observability and controllability are generalizations of the notions of observability and controllability for realizations that were discussed in Section \ref{OCR}.  External observability and controllability are defined for the external behavior $\CC^\FF$ of a fragment, and  generalize  notions of observability and controllability for constraint codes.  Total observability and controllability amount to the combination of both of these properties.

Finally, generalizing results of \cite{FGL12}, we show that the unobservable part of an internally proper  realization $\RR$ lies within its 2-core $\bar{\RR}$, and, dually, that the uncontrollable part of an internally trim  realization $\RR$ lies within $\bar{\RR}$.

\subsection{Observability of fragments}\label{OF}

In general, the term ``observable" applies to systems that have a set $\SSS$ of internal (``state") configurations and a set $\A$ of external variable configurations.  A system is called ``observable" if observation of an external configuration $\ab \in \A$ determines the internal configuration $\sb \in \SSS$.

For a fragment $\FF$ with internal behavior $\Bf^\FF \subseteq \A^{\FF} \times \SSS^{\FF,\mathrm{ext}} \times\SSS^{\FF,\mathrm{int}}$ and external behavior $\CC^\FF = (\Bf^\FF)_{|\A^{\FF} \times \SSS^{\FF,\mathrm{ext}}}$, the definition of observability  depends on whether the  behavior of the system is regarded as $\Bf^\FF$ or $\CC^\FF$, and  on which variables are regarded as internal and which as external.  Consequently, we may define three notions of observability, as follows:
\begin{itemize}
\item A fragment $\FF$ is \emph{externally observable} if the projection $\CC^\FF \to \A^\FF$ is one-to-one;  \ie if  the symbol configuration $\ab$ determines the external state configuration $\sb^{\mathrm{ext}}$.  Evidently $\FF$ is externally observable if and only if the \emph{externally unobservable state configuration space} $\SSS^{\FF,\mathrm{ext},u} = (\CC^\FF)_{:\SSS^{\FF,\mathrm{ext}}}$ is trivial, since $\SSS^{\FF,\mathrm{ext},u}$ is isomorphic to the kernel of this projection.  Alternatively, $\FF$ is externally observable if and only if $|\CC^\FF| = |\bar{\A}^\FF|$, where $\bar{\A}^\FF = (\CC^\FF)_{|\A^\FF}$.
\item A fragment $\FF$ is \emph{internally observable} if the projection $\Bf^\FF \to \CC^\FF$ is one-to-one;  \ie  if the symbol configuration $\ab$ and the external state configuration $\sb^{\mathrm{ext}}$ together determine the internal state configuration $\sb^{\mathrm{int}}$.     Evidently $\FF$ is internally observable if and only if the \emph{internally unobservable state configuration space} $\SSS^{\FF,\mathrm{int},u} =  (\Bf^\FF)_{:\SSS^{\FF,\mathrm{int}}}$ is trivial, since $\SSS^{\FF,\mathrm{int},u}$ is isomorphic to the kernel of this projection.   Alternatively, $\FF$ is internally observable if and only if $|\Bf^\FF| = |\CC^\FF|$, since $\CC^\FF = (\Bf^\FF)_{|\A^{\FF} \times \SSS^{\FF,\mathrm{ext}}}$.
  \item A fragment $\FF$ is \emph{totally observable} if the projection $\Bf^\FF \to \A^\FF$ is one-to-one;  \ie if the symbol configuration $\ab$ determines both $\sb^{\mathrm{ext}}$ and $\sb^{\mathrm{int}}$.      Evidently $\FF$ is totally observable if and only if the \emph{totally unobservable state configuration space} $\SSS^{\FF,\mathrm{tot},u} =  (\Bf^\FF)_{:\SSS^{\FF,\mathrm{ext}} \times \SSS^{\FF,\mathrm{int}}}$ is trivial, since $\SSS^{\FF,\mathrm{tot},u}$ is isomorphic to the kernel of this projection.  Alternatively, $\FF$ is totally observable if and only if $|\Bf^\FF| = |\bar{\A}^\FF|$, since $\bar{\A}^\FF = (\Bf^\FF)_{|\A^\FF}$.    \end{itemize}
  
We note immediately that a fragment  $\FF$ is totally observable if and only if it is both externally and internally observable, since the projection $\Bf^\FF \to \A^\FF$ is the composition of the projections $\Bf^\FF \to \CC^\FF$ and $\CC^\FF \to \A^\FF$.

Evidently $\FF$ is externally observable if and only if its external behavior $\CC^\FF$ is proper at its state configuration space $\SSS^{\FF,\mathrm{ext}}$, in the sense that $(\CC^\FF)_{:\SSS^{\FF,\mathrm{ext}}}$ is trivial.  A leaf fragment $\FF$ is thus externally observable if and only if it is proper at its external state variable.

If $\FF$ has no internal state variables--- \ie if $\FF$ is a constraint code--- then:
 \begin{itemize}
 \item $\FF$ is trivially internally observable, since $\Bf^\FF = \CC^\FF$;
  \item $\FF$ is totally observable if and only if $\FF$ is externally observable.
 \end{itemize}
 Thus for a constraint code $\CC_i \subseteq \A^{(i)} \times \SSS^{(i)}$, the notions of properness at $\SSS^{(i)}$, external observability and total observability coincide.
 
The definition of internal observability of a fragment $\FF$ generalizes the definition of observability of a normal realization $\RR$ in Section \ref{OCR},  because if $\FF$ has no external state variables--- \ie if $\FF$ is a normal realization--- then:
  \begin{itemize}
 \item $\FF$ is trivially externally observable, since $\CC^\FF = \bar{\A}^\FF$;
 \item $\FF$ is internally observable if and only if $\FF$ is observable in the sense of Section \ref{OCR};
 \item $\FF$ is totally observable if and only if $\FF$ is internally observable.
 \end{itemize}
 Thus for a normal realization $\RR$, the notions of observability  in the sense of Section \ref{OCR}, internal observability and total observability coincide.
 
  \vspace{1ex}
  \noindent
  \textbf{Example 1} (trellis fragments, cont.)  Let us see how these definitions apply to a  fragment $\FF^{[j,k)}$ of a conventional state-space (trellis) realization, as shown in Figure \ref{TF}.  
    $\FF^{[j,k)}$ is  externally observable if  the symbol sequence $\ab^{[j,k)}$ determines the ``state transition" $(s_j, s_k)$.  (In \cite{GLF12}, this property is called  ``$[j,k)$-observability.")
    $\FF^{[j,k)}$ is internally observable if the symbol sequence $\ab^{[j,k)}$ and $(s_j, s_k)$ determine the remaining state sequence $\sb^{(j,k)}$.  As we shall show in Section \ref{CFC}, $\FF^{[j,k)}$ is internally observable if it is internally proper;  \ie if all its constraint codes are proper.
 Finally,   $\FF^{[j,k)}$ is totally observable if $\ab^{[j,k)}$ determines the entire state sequence $\sb^{[j,k]}$;  this is what is usually called ``observability" in classical linear systems theory.    Thus if all constraint codes are proper, as we generally  assume, then  $\FF^{[j,k)}$ is totally observable if and only if it is externally observable, so the notions of external, total and classical observability coincide.     \qed

\subsection{Controllability of fragments}\label{CF}
 
  In general, for any of these notions of observability, we will define a corresponding notion of controllability  such that a fragment $\FF$ is controllable if and only if the dual fragment $\FF^\circ$ is observable.  
  
  Since external observability amounts to generalized properness at $\SSS^{\FF,\mathrm{ext}}$, we  define external controllability  as  generalized trimness at $\SSS^{\FF,\mathrm{ext}}$, as follows.  We   say that a linear or group fragment $\FF$ is \emph{externally controllable} if the projection $(\CC^\FF)_{|\SSS^{\FF,\mathrm{ext}}}$ is equal to $\SSS^{\FF,\mathrm{ext}}$ (\ie is surjective, or onto).  Then, by projection/cross-section duality, a fragment $\FF$ is externally controllable if and only if its dual  $\FF^\circ$ is externally observable.
  
  Since internal observability generalizes the notion of observability of realizations, we  define internal controllability to generalize the notion of controllability of realizations, as follows.      We define the \emph{configuration universe} of $\FF$ as  $\U^\FF = \prod_{C(\FF)} \CC_i$, and its \emph{validity space} as $\V^\FF = \{(\ab, \sb^{\mathrm{ext}}, \sb, \sb\} \in \A \times \SSS^{\FF,\mathrm{ext}} \times \SSS^{\FF,\mathrm{int}} \times \SSS^{\FF,\mathrm{int}}\}$;  then its extended behavior is $\bar{\Bf}^\FF =  \U^\FF \cap \V^\FF$.  We   say that $\FF$ is \emph{internally controllable} if the check subspaces $(\U^\FF)^\perp = \prod_{C(\FF^\circ)} (\CC_i)^\perp$ and $(\V^\FF)^\perp = \{(\zerob, \zerob, \hat{\sb}, -\hat{\sb}) \in \hat{\A} \times \hat{\SSS}^{\FF,\mathrm{ext}}\times \hat{\SSS}^{\FF,\mathrm{int}}\times \hat{\SSS}^{\FF,\mathrm{int}}\}$ are independent.  A fragment $\FF$ is then internally controllable if and only if its dual  $\FF^\circ$ is internally observable, because then and only then  the internally unobservable extended dual behavior $(\bar{\Bf}^\circ)^u = (\U^\FF)^\perp \cap (\V^\FF)^\perp$ is trivial.
  
  Finally, a linear or group fragment $\FF$ will be called \emph{totally controllable} if it is both internally and externally controllable.  This happens if and only if its dual $\FF^\circ$ is both internally and externally observable;  \ie if and only if $\FF^\circ$ is totally observable.
  
  Our definitions of internal observability and controllability of a fragment $\FF$ with internal behavior $\Bf^\FF \subseteq \A^\FF \times \SSS^{\FF,\mathrm{ext}} \times \SSS^{\FF,\mathrm{int}}$ are the same as those for a realization $\RR$ with behavior $\Bf \subseteq \A \times \SSS$  if we conflate the symbol and external state variables of $\FF$, so that  $\FF$ may be regarded as a normal realization of its external behavior $\CC^{\FF}$.  Thus we immediately obtain generalizations of all results of Section \ref{OCR}.
  
  In particular, the internally unobservable state configuration space $\SSS^{\FF,\mathrm{int},u}$ of a fragment $\FF$ has been defined as $(\Bf^\FF)_{:\SSS^{\FF,\mathrm{int}}}$.  The \emph{dual internally controllable subspace} of a dual fragment $\FF^\circ$ will  be defined as $\hat{\SSS}^{\FF,\mathrm{int},c} = \{ \hat{\sb} + \hat{\sb}' :  (\hat{\ab}, \hat{\sb}^{\mathrm{ext}}, \hat{\sb}, \hat{\sb}') \in (\U^\FF)^\perp\}$,  namely the set of syndromes $\hat{\sb} + \hat{\sb}' \in \hat{\SSS}^{\FF,\mathrm{ext}}$ that occur as $(\hat{\ab}, \hat{\sb}^\mathrm{ext}, \hat{\sb}, \hat{\sb}')$ runs through $(\U^\FF)^\perp$.  Similarly,  the \emph{internally controllable subspace} of $\FF$ will be defined as $\SSS^{\FF,\mathrm{int},c} = \{\sb - \sb' :  (\ab, \sb^{\mathrm{ext}}, \sb, \sb') \in \U^\FF\}$.  Then, as in Section \ref{OCR}, we have:

\vspace{1ex}
\noindent
\textbf{Theorem} (\textbf{internal unobservability/controllability duality}). The internally unobservable state configuration space $\SSS^{\FF,\mathrm{int},u}$ of a linear or group fragment $\FF$ and the dual internally controllable subspace $\hat{\SSS}^{\FF,\mathrm{int},c}$ of its dual fragment $\FF^\circ$ are orthogonal;  \ie  $\hat{\SSS}^{\FF,\mathrm{int},c} = (\SSS^{\FF,\mathrm{int},u})^\perp$.  Similarly, $\SSS^{\FF,\mathrm{int},c} = (\hat{\SSS}^{\FF,\mathrm{int},u})^\perp$. Thus $\FF$ (resp.\ $\FF^\circ$) is internally controllable if and only if $\SSS^{\FF,\mathrm{int},c} = \SSS^{\FF,\mathrm{int}}$ (resp.\ $\hat{\SSS}^{\FF,\mathrm{int},c} = \hat{\SSS}^{\FF,\mathrm{int}}$).  \qed \vspace{1ex}

\noindent
\textbf{Theorem} (\textbf{internal controllability test}).  For a linear or group fragment $\FF$ with extended behavior $\bar{\Bf}^\FF \subseteq \U^\FF$ and internally controllable subspace $\SSS^{\FF,\mathrm{int},c} \subseteq \SSS^{\FF,\mathrm{int}}$, we have $|\U^\FF|/|\bar{\Bf}^\FF| = |\SSS^{\FF,\mathrm{int},c}| \le |\SSS^{\FF,\mathrm{int}}|$, or in the linear case $\dim \U^\FF - \dim \bar{\Bf}^\FF = \dim \SSS^{\FF,\mathrm{int},c} \le \dim \SSS^{\FF,\mathrm{int}}$,
with equality if and only if $\FF$ is internally controllable.  \qed \vspace{1ex}

  \noindent
  \textbf{Example 1} (trellis fragments, cont.)  Again, let us see how these definitions apply to a  fragment $\FF^{[j,k)}$ of a conventional state-space realization as  in Figure \ref{TF}. 
  $\FF^{[j,k)}$ is externally controllable (or ``$[j,k)$-controllable" \cite{GLF12}) if all state transitions  $(s_j, s_k)$ can occur.  This is what is usually called ``state controllability" (or ``reachability") in classical linear systems theory.   
We shall show shortly that $\FF^{[j,k)}$ is internally controllable if it is  internally trim, so in this case external, total and classical state controllability  coincide.   \qed 

\subsection{Behavioral controllability and observability}

In this section, we   compare and contrast our notion of external controllability to Willems' proposed generalization  \cite[Fig.\ 14]{W07} of behavioral controllability to $n$-dimensional ($n$-D) systems.  We also consider the dual notions of observability.

Willems' notion of behavioral controllability is illustrated in Figure \ref{WSS}.  Here $\FF$ and $\FF'$ are disjoint fragments of a realization $\RR$ of a  set $\CC$ of trajectories, and $\FF''$ represents the remainder of the realization.  The realization is said to be \emph{behaviorally controllable} with respect to fragments $\FF, \FF'$ if $\CC_{|\A^\FF \times \A^{\FF'}} = \CC_{|\A^\FF} \times \CC_{|\A^{\FF'}}$;  \ie if for any two valid partial trajectories $\ab^\FF \in \CC_{|\A^\FF}, \ab^{\FF'} \in \CC_{|\A^{\FF'}}$, there is a trajectory $\ab \in \CC$ whose projection onto $\A^\FF \times \A^{\FF'}$ is $(\ab^\FF, \ab^{\FF'})$.

\begin{figure}[h]
\setlength{\unitlength}{5pt}
\centering
\begin{picture}(30,9)(5, 3)
\put(18,2.5){\framebox(5,5){$\CC^{\FF''}$}}
\put(31,2.5){\framebox(5,5){$\CC^{\FF'}$}}
\multiput(23,5)(30,0){1}{\line(1,0){8}}
\multiput(10,5)(20,0){1}{\line(1,0){8}}
\multiput(5,2.5)(40,0){1}{\framebox(5,5){$\CC^\FF$}}
\multiput(7.5,7.5)(26,0){2}{\line(0,1){3}}
\multiput(6,10.5)(26,0){2}{\line(1,0){3}}
\put(6,11){$\A^\FF$}
\put(19,11){$\A^{\FF''}$}
\multiput(20.5,7.5)(20,0){1}{\line(0,1){3}}
\multiput(19,10.5)(20,0){1}{\line(1,0){3}}
\put(32,11){$\A^{\FF'}$}
\put(11,6){$\SSS^{\FF,\mathrm{ext}}$}
\put(24,6){$\SSS^{\FF',\mathrm{ext}}$}
\end{picture}
\caption{State realization $\RR$ with two disjoint fragments $\FF, \FF'$.}
\label{WSS}
\end{figure}

For trim 1-D state realizations and two fragments consisting of a ``past" up to time $t$ and a ``future" from some time $t'$ on, behavioral controllability  is equivalent to external controllability, since  both definitions require that any trajectory up to time $t$ (\ie any state $s(t)$) can be connected by a valid path during $[t, t')$ to any trajectory from time $t'$ on (\ie any state $s(t')$).

\pagebreak
Let us introduce the  trimmed external state variables $\bar{\SSS}^\FF = (\CC^\FF)_{|\SSS^{\FF,\mathrm{ext}}}$ and $\bar{\SSS}^{\FF'} = (\CC^{\FF'})_{|\SSS^{\FF',\mathrm{ext}}}$.  We now observe that:

\vspace{1ex}
\noindent
\textbf{Theorem} (\textbf{behavioral controllability})  The state realization $\RR$ of Figure \ref{WSS} is behaviorally controllable if and only if 
$$(\CC^{\FF''})_{|\bar{\SSS}^{\FF}\times \bar{\SSS}^{\FF'}} = \bar{\SSS}^\FF \times \bar{\SSS}^{\FF'}.$$

\vspace{1ex}
\noindent
\textsc{Proof}:  $\RR$ is behaviorally controllable if and only if for every  $\sb^\FF \in  \bar{\SSS}^{\FF}$ and $\sb^{\FF'} \in  \bar{\SSS}^{\FF'}$ we have $(\sb^\FF, \sb^{\FF'}) \in  (\CC^{\FF''})_{|\SSS^{\FF,\mathrm{ext}}\times \SSS^{\FF',\mathrm{ext}}}$;  \ie $\bar{\SSS}^{\FF}\times \bar{\SSS}^{\FF'} \subseteq (\CC^{\FF''})_{|\SSS^{\FF,\mathrm{ext}}\times \SSS^{\FF',\mathrm{ext}}}$.  But this is true if and only if $\bar{\SSS}^\FF \times \bar{\SSS}^{\FF'} \subseteq (\CC^{\FF''})_{|\bar{\SSS}^{\FF}\times \bar{\SSS}^{\FF'}}$, whereas
 it is always true that $(\CC^{\FF''})_{|\bar{\SSS}^{\FF}\times \bar{\SSS}^{{\FF}'}} \subseteq \bar{\SSS}^\FF \times \bar{\SSS}^{\FF'}$. \qed \vspace{1ex}
 
Thus behavioral controllability is  a kind of memorylessness, somewhat reminiscent of  marginal independence in probability theory.

\begin{figure}[h]
\setlength{\unitlength}{5pt}
\centering
\begin{picture}(30,9)(5, 3)
\put(18,2.5){\framebox(5,5){$\CC^{\FF''}$}}
\put(43,2.5){\framebox(5,5){$\bar{\CC}^{\FF'}$}}
\multiput(23,5)(30,0){1}{\line(1,0){8}}
\multiput(-1,5)(20,0){1}{\line(1,0){7}}
\multiput(-6,2.5)(40,0){1}{\framebox(5,5){$\bar{\CC}^\FF$}}
\multiput(-3.5,7.5)(49,0){2}{\line(0,1){3}}
\multiput(-5,10.5)(49,0){2}{\line(1,0){3}}
\put(-5,11){$\A^\FF$}
\put(1,6){$\bar{\SSS}^{\FF}$}
\multiput(6,4)(50,0){1}{\line(0,1){2}}
\put(7,4.5){$\hookrightarrow$}
\multiput(10,3)(50,0){1}{\line(-4,1){4}}
\multiput(10,7)(50,0){1}{\line(-4,-1){4}}
\multiput(10,3)(50,0){1}{\line(0,1){4}}
\multiput(10,5)(30,0){1}{\line(1,0){8}}
\put(19,11){$\A^{\FF''}$}
\multiput(20.5,7.5)(20,0){1}{\line(0,1){3}}
\multiput(19,10.5)(20,0){1}{\line(1,0){3}}
\put(44,11){$\A^{\FF'}$}
\put(11,6){$\SSS^{\FF,\mathrm{ext}}$}
\put(24,6){$\SSS^{\FF',\mathrm{ext}}$}
\put(39,6){$\bar{\SSS}^{\FF'}$}
\multiput(35,4)(50,0){1}{\line(0,1){2}}
\put(31.5,4.5){$\hookleftarrow$}
\multiput(31,3)(50,0){1}{\line(4,1){4}}
\multiput(31,7)(50,0){1}{\line(4,-1){4}}
\multiput(31,3)(50,0){1}{\line(0,1){4}}
\multiput(35,5)(30,0){1}{\line(1,0){8}}
\put(5,1){\dashbox(31,8){}}
\put(34,10){$\bar{\FF}''$}
\end{picture}
\caption{Equivalent reduced state realization.}
\label{WSSp}
\end{figure}
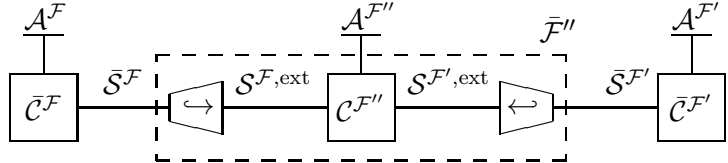

Figure \ref{WSSp} expands Figure \ref{WSS} to show the trimmed state variables $\bar{\SSS}^\FF$ and $\bar{\SSS}^{\FF'}$, and defines a reduced fragment $\bar{\FF}''$ involving them.  We observe that $\RR$ is behaviorally controllable if and only if the reduced fragment $\bar{\FF}''$ is externally controllable;  \ie $(\CC^{\FF''})_{|\bar{\SSS}^{\FF}\times \bar{\SSS}^{\FF'}} = \bar{\SSS}^\FF \times  \bar{\SSS}^{\FF'}$.  

  If $\CC^\FF$ and $\CC^{\FF'}$ are trim at $\SSS^{\FF,\mathrm{ext}}$ and $\SSS^{\FF',\mathrm{ext}}$, respectively, as we may assume when $\SSS^{\FF,\mathrm{ext}}$ and $\SSS^{\FF',\mathrm{ext}}$ are single state variables,  then $\bar{\SSS}^{\FF} = \SSS^{\FF,\mathrm{ext}}$ and $\bar{\SSS}^{\FF'} = \SSS^{\FF',\mathrm{ext}}$, so $\bar{\FF}''$ is externally controllable if and only if $\FF''$ is.
    However, if $\SSS^{\FF,\mathrm{ext}}$ and $\SSS^{\FF',\mathrm{ext}}$ represent multiple state variables, then in general we cannot expect $\CC^\FF$ and $\CC^{\FF'}$ to be trim at $\SSS^{\FF,\mathrm{ext}}$ and $\SSS^{\FF',\mathrm{ext}}$, so while $\bar{\FF}''$ is externally controllable if  $\FF''$ is so,  the converse may not hold.  
      
      In summary, our definition of external controllability is equivalent to Willems' notion of behavioral controllability under natural trimness conditions.  The main difference is that Willems focusses on the leaf fragments $\FF$ and $\FF'$, whereas we focus on the central fragment $\FF''$. 
  
  Dually, we may define a realization $\RR$ with two disjoint fragments $\FF, \FF'$ as in Figure \ref{WSS} as \emph{behaviorally observable} w.r.t.\ $\FF, \FF'$ if $\CC_{:\A^\FF \times \A^{\FF'}} = \CC_{:\A^\FF} \times \CC_{:\A^{\FF'}}$;  \ie if whenever $(\ab^\FF, \zerob^{\FF''}, \ab^{\FF'}) \in \CC$, then $(\ab^\FF, \zerob^{\FF''}, \zerob^{\FF'}) \in \CC$ and $(\zerob^\FF, \zerob^{\FF''}, \ab^{\FF'}) \in \CC$.\footnote{In the  set-theoretic setting of \cite{W07}, we must take cross-sections with respect to every  $\ab^{\bar{\FF}} \in (\CC^{\bar{\FF}})_{|\A^{\bar{\FF}}}$, not just $\zerob^{\bar{\FF}}$.}
    For proper 1-D state realizations and  fragments $\FF$ and $\FF'$ defined on $(-\infty,t)$ and  $[t',\infty)$, respectively, behavioral observability  is equivalent to external observability, since  both  are true if and only if every trajectory $(\ab^{(-\infty,t)}, \zerob^{[t,t')}, \ab^{[t',\infty)}) \in \CC$  passes through the zero state at times $t$ and $t'$.
    
    In general, it is easy to show that $\RR$  is behaviorally observable if and only if 
$(\CC^{\FF''})_{:\underline{\SSS}^{\FF}\times \underline{\SSS}^{\FF'}} = \underline{\SSS}^\FF \times \underline{\SSS}^{\FF'},$
where $\underline{\SSS}^\FF = (\CC^\FF)_{:\SSS^{\FF,\mathrm{ext}}}$ and  $\underline{\SSS}^{\FF'} = (\CC^\FF)_{:\SSS^{\FF',\mathrm{ext}}}$.   Thus behavioral observability is another kind of memorylessness,  somewhat reminiscent of  conditional independence in probability theory.

      If $\CC^\FF$ and $\CC^{\FF'}$ are proper at $\SSS^{\FF,\mathrm{ext}}$ and $\SSS^{\FF',\mathrm{ext}}$, respectively--- \ie if $\ab^\FF = \zerob$ implies $\sb^\FF = \zerob$, and similarly for $\FF'$--- then $\underline{\SSS}^\FF = \{\zerob\}$ and $\underline{\SSS}^{\FF'} = \{\zerob\}$, so this is equivalent to the requirement that $\FF''$ be externally observable;  \ie that $(\CC^{\FF''})_{:\SSS^{\FF,\mathrm{ext}}\times \SSS^{\FF',\mathrm{ext}}} = \{\zerob\} \times \{\zerob\}$.  

  \subsection{Connecting fragments, continued}\label{CFC}
  
We now continue our study of connected fragments, begun in Section \ref{CF1}.  Again, we connect a pair of fragments $\FF_1$ and $\FF_2$ with isomorphic external state spaces $\SSS_j$ and $\SSS_j'$ via an isomorphism $\varphi: \SSS_j \to \SSS_j'$ to form a combined fragment $\FF_{12}$  as in Figure \ref{linkCp}.   However, we now include symbol configuration spaces $\A^{(1)}$ and $\A^{(2)}$, as shown in Figure \ref{linkCpp}.

\begin{figure}[h]
\setlength{\unitlength}{5pt}
\centering
\begin{picture}(30,7)(-5, 3)
\put(-17,7){\line(1,0){8}}
\put(-17,6){\line(0,1){2}}
\put(-16.5,8){$\A^{(1)}$}
\put(-17,3){\line(1,0){8}}
\put(-16.5,4){$\SSS^{(1 \setminus j)}$}
\multiput(-3,5)(50,0){1}{\line(1,0){8}}
\multiput(8,5)(30,0){1}{\line(1,0){8}}
\put(4.5,9.5){$\CC_{12}$}
\put(-10,1.5){\dashbox(33,7){}}
\multiput(-9,2.5)(40,0){1}{\framebox(6,5){$\CC_1$}}
\multiput(5,3.5)(40,0){1}{\framebox(3,3){$\lra$}}
\put(16,2.5){\framebox(6,5){$\CC_{2}$}}
\put(0,6){$\SSS_j$}
\put(11,6){$\SSS'_j$}
\put(22,7){\line(1,0){8}}
\put(30,6){\line(0,1){2}}
\put(24,8){$\A^{(2)}$}
\put(22,3){\line(1,0){8}}
\put(24,4){$\SSS^{(2\setminus j)}$}
\end{picture}
\caption{Connecting two fragments via an isomorphism constraint $\SSS_j \lra \SSS'_j$.}
\label{linkCpp}
\end{figure}

We may now extend the connected fragment lemma of Section \ref{CF1} as follows:

\vspace{1ex}
\noindent
\textbf{Lemma} (\textbf{connected fragments, cont.})  If two linear or group fragments $\FF_1, \FF_2$  are connected via an isomorphism between  state spaces $\SSS_j$ and $\SSS_j'$, and $\FF_{12}$ is the combined fragment, then: \\
\indent (a)   If $\FF_1$ and $\FF_2$ are  trim, then $\FF_{12}$ is  trim. \\
\indent (b)  If $\FF_1$ and $\FF_2$ are  proper, then $\FF_{12}$ is  proper. \\
\indent (c)  If $\FF_1$ and $\FF_2$ are  externally observable, then $\FF_{12}$ is  externally observable.  \\
\indent (d) If $\FF_1$ and $\FF_2$ are  externally controllable, then $\FF_{12}$ is  externally controllable.  \\
\indent (e)  If $\FF_1$ and $\FF_2$ are proper and internally observable, then $\FF_{12}$ is internally observable. \\
\indent (f)  If $\FF_1$ and $\FF_2$ are trim and internally controllable, then $\FF_{12}$ is internally controllable.  \\
\indent (g) If $\FF_1$ and $\FF_2$ are proper and totally observable, then $\FF_{12}$ is totally observable. \\
\indent (h) If $\FF_1$ and $\FF_2$ are trim and totally controllable, then $\FF_{12}$ is totally controllable.

\vspace{1ex}
\noindent
\textsc{Proof}:  (a)--(b)  were proved in Section \ref{CF1}.

(c)   If $\FF_1$ is  externally observable, then $\ab^{(1)} = \zerob$ implies $(\sb^{(1\setminus j)}, s_j) = (\zerob, 0)$, and similarly for $\FF_2$.  Hence $\ab^{(12)} = (\ab^{(1)}, \ab^{(2)}) = (\zerob, \zerob)$ implies $\sb^{(12)} = (\sb^{(1\setminus j)}, \sb^{(2 \setminus j)}) = (\zerob, \zerob)$, so $\FF_{12}$ is  externally observable. 
(d)  follows  from (c) by observability/controllability duality. 

(e) If $\FF_1$ is proper, then $\ab^{(1)} = \zerob$ and $\sb^{(1 \setminus j)} = \zerob$ imply $s_j = 0$, and similarly for $\FF_2$.  If $\FF_1$ is internally observable, then this implies $\sb^{(1),\mathrm{int}} = \zerob$, and similarly for $\FF_2$. Hence $(\ab^{(12)}, \sb^{(12)}) = (\zerob, \zerob)$ implies $\sb^{(12),\mathrm{int}} = \zerob$, so $\FF_{12}$ is internally observable. 
(f)  follows  from (e) by trim/proper  and  observability/controllability duality.

(g) Since a fragment is totally observable if and only if it is internally and externally observable, it follows from (c) and (e) that if $\FF_1$ and $\FF_2$ are proper and totally observable, then $\FF_{12}$ is totally observable. (h)  follows  from (g) by trim/proper  and observability/controllability duality. \qed \vspace{1ex} 

Since constraint codes are trivially internally observable and controllable, we may now  extend the cycle-free fragment theorem of Section \ref{CF1} as follows:

\vspace{1ex}
\noindent
\textbf{Theorem} (\textbf{cycle-free fragments, cont.})  For a cycle-free linear or group fragment $\FF$: \\
\indent (a) if $\FF$ is internally  proper, then $\FF$ is   proper and internally observable. \\
\indent (b) if $\FF$ is internally  trim, then $\FF$ is   trim and internally controllable. \qed

\vspace{1ex}
  \noindent
  \textbf{Example 1} (trellis fragments, cont.)  A trellis fragment is cycle-free;  thus if all its constraint codes are proper (``instantaneously invertible"), then it is internally observable, and if all are trim, then it is internally controllable. \qed \vspace{1ex}
  
More concretely, we can sketch an iterative algorithm for determining the internal state values of an internally proper cycle-free fragment $\FF$, given the external symbol configuration $\ab$, as follows.  A cycle-free fragment has at least two leaf constraints.  Given a proper leaf constraint $\CC_i$ and its symbol configuration $\ab^{(i)}$, the single external state value $s_j$ of that leaf constraint may be determined.  The leaf constraint may then be  stripped from $\FF$, leaving a smaller internally proper cycle-free fragment $\FF'$.  This procedure may be iterated until all internal state values have been determined.  
  
  Finally, as a corollary, we  extend the result of \cite[Theorem 11]{FGL12}  to cyclic realizations, as follows:
  
\vspace{1ex}
\noindent
\textbf{Theorem} (\textbf{unobservable/uncontrollable realizations and cycles}) \\
\indent (a) An internally proper cycle-free linear or group realization $\RR$ is internally observable.  \\
\indent (b) An internally proper cyclic linear or group realization $\RR$ is internally observable if and only if its 2-core $\bar{\RR}$ is internally observable. \\
\indent (c) An internally trim cycle-free linear or group realization $\RR$ is  internally controllable. \\
\indent (d) An internally trim cyclic linear or group realization $\RR$ is internally controllable if and only if its 2-core $\bar{\RR}$ is internally controllable.

  \vspace{1ex}
  \noindent
  \textsc{Proof}:  (a) follows from the cycle-free fragment theorem.  For (b), we observe that since a proper cycle-free leaf fragment is internally observable, the combination of such a fragment with an internally proper 2-core is internally observable if the 2-core is internally observable, from part (e) of the connected fragment theorem;  on the other hand, if the 2-core is not internally observable, then it supports an unobservable sequence, so the combination supports an unobservable sequence.  Parts (c) and (d) follow from trim/proper and controllability/observability duality. \qed

\subsection{State-trimness}

A realization $\RR$ with behavior $\Bf$ is said to be \emph{state-trim} at $\SSS_j$ if $\Bf_{|\SSS_j} = \SSS_j$. We have seen that an internally trim cycle-free realization is state-trim at all its internal state variables.  In this section we will consider  state-trimness in  cyclic realizations.

Let $\RR$ be a cyclic normal realization $\RR$ with internal behavior $\Bf$ and external behavior $\CC$.  We will suppose that 
the fragment $\RR^{(\setminus j)}$ that results from cutting one edge $\SSS_j$ is connected--- \ie the edge $\SSS_j$ is not a cut set.  This can happen only when $\RR$ is cyclic.
Then $\RR^{(\setminus j)}$ is a  trellis fragment that has two external state variables with values $s_j \in \SSS_j$ and $s'_j \in \SSS_j$,  symbol configurations $\ab \in \A$, and  external behavior $\CC^{(\setminus j)} \subseteq \A \times \SSS_j \times \SSS_j$.  
The original realization $\RR$ may be recovered by imposing an equality constraint  
on $s_j$ and $s'_j$.

Thus we will regard $\RR$  as a realization with a single constraint code $\CC^{(\setminus j)}$ and  internal state space $\SSS_j$.  Its extended internal behavior is $\bar{\Bf} = \CC^{(\setminus j)} \cap \V$, where $\V = \A \times \CC_{=\SSS_j}$ is its validity space, and its behavior is $\Bf = \bar{\Bf}_{|\A \times \SSS_j}$.

In this context, $\RR$ is  observable if and only if $(\SSS_j)^u = \Bf_{:\SSS_j}$
is trivial.  Dually, $\RR$ is  controllable if and only if $(\SSS_j)^c = ((\Bf^\circ)^\perp)_{|\SSS_j} = (\CC^{(\setminus j)} + (\V^\circ)^\perp)_{|\SSS_j}$
is equal to $\SSS_j$, where $\CC_{\sim \SSS_j} = \{(s_j, -s'_j) \in \SSS_j \times \SSS_j\}$ and $(\V^\circ)^\perp = \{\zerob\} \times \CC_{\sim \SSS_j}$.   
Figure \ref{OCDx} shows realizations of $(\SSS_j)^u$, $(\SSS_j)^c$ and  their duals $(\hat{\SSS}_j)^c$,  $(\hat{\SSS}_j)^u$.

\begin{figure}[h]
\setlength{\unitlength}{5pt}
\centering
\begin{picture}(64,16)(-2, 2)
\put(0,15){\line(1,0){4}}
\put(-1.5,14.3){$\square$}
\put(1,16){$\A$}
\multiput(4,12.5)(40,0){1}{\framebox(6,5){$\CC^{(\setminus j)}$}}
\put(10,13){\line(1,0){9}}
\put(10,17){\line(1,0){9}}
\put(11,17.5){$s_j \in \SSS_j$}
\put(19,17){\line(0,-1){1}}
\put(11,13.5){$s_j' \in \SSS_j$}
\put(19,13){\line(0,1){1}}
\multiput(18,14)(40,0){1}{\framebox(2,2){$=$}}
\put(20,15){\line(1,0){5}}
\put(21,15.5){$\SSS_j$}
\put(6,10){(a)}

\put(31.5,14.3){$\blacksquare$}
\put(33,15){\line(1,0){4}}
\put(34,16){$\hat{\A}$}
\multiput(37,12.5)(40,0){1}{\framebox(8,5){$(\CC^{(\setminus j)})^\perp$}}
\put(45,13){\line(1,0){9}}
\put(45,17){\line(1,0){9}}
\put(46,17.5){$\hat{s}_j \in \hat{\SSS}_j$}
\put(46,13.5){$\hat{s}_j' \in \hat{\SSS}_j$}
\put(53.5,16){$\circ$}
\multiput(53,14)(40,0){1}{\framebox(2,2){$+$}}
\put(53.5,13.0){$\circ$}
\put(55,15){\line(1,0){5}}
\put(56,15.5){$\hat{\SSS}_j$}
\put(41,10){(c)}

\put(-1.5,4.3){$\blacksquare$}
\put(0,5){\line(1,0){4}}
\put(1,6){$\A$}
\multiput(4,2.5)(40,0){1}{\framebox(6,5){$\CC^{(\setminus j)}$}}
\put(10,3){\line(1,0){9}}
\put(10,7){\line(1,0){9}}
\put(11,7.5){$s_j \in \SSS_j$}
\put(54,7){\line(0,-1){1}}
\put(11,3.5){$s_j' \in \SSS_j$}
\put(19,3){\line(0,1){1}}
\multiput(18,4)(40,0){1}{\framebox(2,2){$+$}}
\put(20,5){\line(1,0){5}}
\put(21,5.5){$\SSS_j$}
\put(6,0){(b)}

\put(31.5,4.3){$\square$}
\put(33,5){\line(1,0){4}}
\put(34,6){$\hat{\A}$}
\multiput(37,2.5)(40,0){1}{\framebox(8,5){$(\CC^{(\setminus j)})^\perp$}}
\put(45,3){\line(1,0){9}}
\put(45,7){\line(1,0){9}}
\put(46,7.5){$\hat{s}_j \in \hat{\SSS}_j$}
\put(46,3.5){$\hat{s}_j' \in \hat{\SSS}_j$}
\put(18.5,6){$\circ$}
\multiput(53,4)(40,0){1}{\framebox(2,2){$=$}}
\put(53.5,3.0){$\circ$}
\put(55,5){\line(1,0){5}}
\put(56,5.5){$\hat{\SSS}_j$}
\put(41,0){(d)}
\end{picture}
\caption{Dual normal realizations of (a) $(\SSS_j)^u$; (b) $(\SSS_j)^c$; (c) $(\hat{\SSS}_j)^c$; (d) $(\hat{\SSS}_j)^u$.}
\label{OCDx}
\end{figure}

\begin{figure}[h]
\setlength{\unitlength}{5pt}
\centering
\begin{picture}(64,16)(-2, 2)
\put(0,15){\line(1,0){4}}
\put(-1.5,14.3){$\blacksquare$}
\put(1,16){$\A$}
\multiput(4,12.5)(40,0){1}{\framebox(6,5){$\CC^{(\setminus j)}$}}
\put(10,13){\line(1,0){9}}
\put(10,17){\line(1,0){9}}
\put(11,17.5){$s_j \in \SSS_j$}
\put(19,17){\line(0,-1){1}}
\put(11,13.5){$s_j' \in \SSS_j$}
\put(19,13){\line(0,1){1}}
\multiput(18,14)(40,0){1}{\framebox(2,2){$=$}}
\put(20,15){\line(1,0){5}}
\put(21,15.5){$\SSS_j$}
\put(6,10){(a)}

\put(31.5,14.3){$\square$}
\put(33,15){\line(1,0){4}}
\put(34,16){$\hat{\A}$}
\multiput(37,12.5)(40,0){1}{\framebox(8,5){$(\CC^{(\setminus j)})^\perp$}}
\put(45,13){\line(1,0){9}}
\put(45,17){\line(1,0){9}}
\put(46,17.5){$\hat{s}_j \in \hat{\SSS}_j$}
\put(46,13.5){$\hat{s}_j' \in \hat{\SSS}_j$}
\put(53.5,16){$\circ$}
\multiput(53,14)(40,0){1}{\framebox(2,2){$+$}}
\put(53.5,13.0){$\circ$}
\put(55,15){\line(1,0){5}}
\put(56,15.5){$\hat{\SSS}_j$}
\put(41,10){(c)}

\put(-1.5,4.3){$\square$}
\put(0,5){\line(1,0){4}}
\put(1,6){$\A$}
\multiput(4,2.5)(40,0){1}{\framebox(6,5){$\CC^{(\setminus j)}$}}
\put(10,3){\line(1,0){9}}
\put(10,7){\line(1,0){9}}
\put(11,7.5){$s_j \in \SSS_j$}
\put(54,7){\line(0,-1){1}}
\put(11,3.5){$s_j' \in \SSS_j$}
\put(19,3){\line(0,1){1}}
\multiput(18,4)(40,0){1}{\framebox(2,2){$+$}}
\put(20,5){\line(1,0){5}}
\put(21,5.5){$\SSS_j$}
\put(6,0){(b)}

\put(31.5,4.3){$\blacksquare$}
\put(33,5){\line(1,0){4}}
\put(34,6){$\hat{\A}$}
\multiput(37,2.5)(40,0){1}{\framebox(8,5){$(\CC^{(\setminus j)})^\perp$}}
\put(45,3){\line(1,0){9}}
\put(45,7){\line(1,0){9}}
\put(46,7.5){$\hat{s}_j \in \hat{\SSS}_j$}
\put(46,3.5){$\hat{s}_j' \in \hat{\SSS}_j$}
\put(18.5,6){$\circ$}
\multiput(53,4)(40,0){1}{\framebox(2,2){$=$}}
\put(53.5,3.0){$\circ$}
\put(55,5){\line(1,0){5}}
\put(56,5.5){$\hat{\SSS}_j$}
\put(41,0){(d)}
\end{picture}
\caption{Dual normal realizations of (a) $\bar{\SSS}_j$; (b) $\underline{\SSS}_j$; (c) $\underline{\hat{\SSS}}_j$; (d) $\bar{\hat{\SSS}}_j$.}
\label{OCDa}
\end{figure}
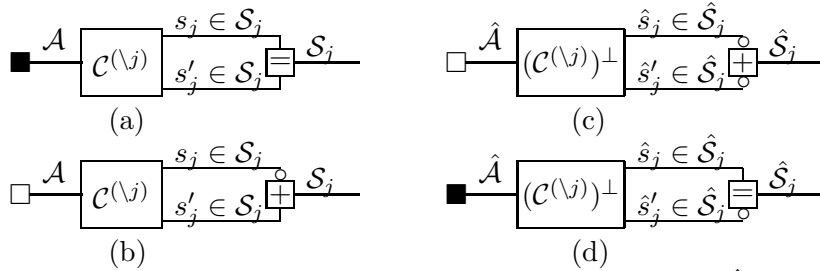

Figure \ref{OCDa}(a) shows a realization of the \emph{trimmed state space} $\bar{\SSS}_j = \Bf_{|\SSS_j}$.    
Evidently $\RR$ is  state-trim at $\SSS_j$ if and only if $\bar{\SSS}_j = \SSS_j$.  

The dual realization to Figure \ref{OCDa}(a), shown in Figure \ref{OCDa}(c), realizes the dual space $\underline{\hat{\SSS}}_j =  \{\hat{s}_j + \hat{s}_j' : (\zerob, \hat{s}_j, \hat{s}_j') \in (\CC^{(\setminus j)})^\perp\}$.   Since $\underline{\hat{\SSS}}_j = (\bar{\SSS}_j)^\perp$, $\RR$ is state-trim at $\SSS_j$ if and only if $\underline{\hat{\SSS}}_j$ is trivial. 

Similarly, Figure \ref{OCDa}(d) realizes the  trimmed state space $\bar{\hat{\SSS}}_j = (\Bf^\circ)_{|\hat{\SSS}_j}$,
and Figure \ref{OCDa}(b) realizes its dual space $\underline{\SSS}_j =  \{s_j - s_j' : (\zerob, s_j, s_j') \in \CC^{(\setminus j)}\} = (\bar{\hat{\SSS}}_j)^\perp$.  $\RR^\circ$ is state-trim at $\hat{\SSS}_j$ if and only if $\bar{\hat{\SSS}}_j = \hat{\SSS}_j$, or if and only if
$\underline{\SSS}_j$ is trivial. 

We will say that $\RR$ is \emph{dual state-trim} at $\SSS_j$ if $\RR^\circ$ is state-trim at $\hat{\SSS}_j$;  \ie if $\underline{\SSS}_j$ is trivial, or if $s_j' = s_j$ for all $(\zerob, s_j, s_j') \in \CC^{(\setminus j)}$.    Thus if we define the \emph{unobservable transition space} $\U^{(\setminus j)} = (\CC^{(\setminus j)})_{:\SSS_j \times \SSS_j} = \{(s_j, s_j') \in \SSS_j \times \SSS_j : (\zerob, s_j, s_j') \in \CC^{(\setminus j)}\}$, as in \cite{GLF12}, then:   \vspace{-1ex}
\begin{itemize}
\item $\RR$ is dual state-trim at $\SSS_j$ if and only if $\U^{(\setminus j)}$ is diagonal;  \ie $s_j' = s_j$ for all $(s_j, s_j') \in \U^{(\setminus j)}$; \vspace{-1ex}
\item $\RR$ is observable if and only if $\U^{(\setminus j)}$ has no diagonal elements $(s_j, s_j)$ other than $(0,0)$; \vspace{-1ex}
\item $\RR^{(\setminus j)}$ is externally observable if and only if $\U^{(\setminus j)}$ is trivial.
\end{itemize}
\vspace{-1ex}

Consequently we obtain  the following theorem, which generalizes \cite[Theorem 5.2]{GLF12}:

\vspace{1ex}
\noindent
\textbf{Theorem} (\textbf{state-trimness}). Let  $\RR^{(\setminus j)}$ be a connected fragment of a normal linear or group realization $\RR$ that results from cutting an edge $\SSS_j$.  
\vspace{1ex}

\indent (a)  $\RR^{(\setminus j)}$ is externally observable if and only if $\RR$ is both dual state-trim at $\SSS_j$ and observable. \\
\indent (b) $\RR^{(\setminus j)}$ is externally controllable if and only if $\RR$ is both  state-trim at $\SSS_j$ and controllable.

\vspace{1ex}
\noindent
\textsc{Proof}:  (a) $\U^{(\setminus j)}$ is trivial if and only if all elements of $\U^{(\setminus j)}$ are diagonal and $\U^{(\setminus j)}$ has no diagonal elements other than $(0,0)$. \\
\indent (b)  From (a), by state-trim duality and observability/controllability duality.  \qed 
\vspace{1ex}

A realization that is not state-trim may be made so by the local reduction of state-trimming.  Therefore we may assume that all realizations $\RR$  are state-trim and dual state-trim everywhere.

Given state-trimness, this theorem  shows that if $\RR$ is  controllable, then every connected fragment $\RR^{(\setminus j)}$ that results from cutting an edge $\SSS_j$ is externally controllable;  \ie all transitions $(s_j, s_j') \in \SSS_j \times \SSS_j$ are possible.  Dually, assuming dual state-trimness, every fragment $\RR^{(\setminus j)}$ of  an  observable realization $\RR$ is externally observable.

\section{Conclusion}\label{Section 8}

This paper develops some fundamental properties of linear and group codes on general graphs, using only elementary group and graph theory, including elementary group duality theory.

Remarkably, these tools suffice to develop a structure theory for realizations on cycle-free graphs.  For cyclic graphs, on the other hand, while the results of this paper may be a good starting point,  there remain many open questions.

Our decomposition results show that, with little loss of generality, we can focus on realizations made up of trim and proper  constraints, plus interface nodes.  For a cyclic realization, we can focus on its 2-core.  

Moreover,  we recall that Internal Node Theorem of \cite[Theorem 10]{F03} shows that for any group or linear realization $\RR$ with maximum constraint code size $|\CC_i|_{\max}$, there exists an equivalent realization $\RR'$ in which all constraints have degree  $\le 3$ and the maximum constraint code size is upperbounded by $|\CC_i|_{\max}$.  Consequently, we can probably focus on realizations with degree-3 (cubic) constraints.
 
On the other hand,  our results so far say little about the properties of symbol configurations, which ultimately determine  properties of minimal realizations.  We have not  yet arrived at  the well-known ``shortest basis theorem" \cite{V99, F11} for minimal  conventional trellis realizations, much less the related results of Koetter and Vardy \cite{KV03} and subsequent authors (\eg \cite{GLW11, GLW11b, GLF12, CB14}) for tail-biting trellis realizations.  The theory of this paper should be extended to cover such results, as well as generalizations of controller and observer granules as in \cite{FT93, FT04}.

\section*{Acknowledgments}

I thank Heide Gluesing-Luerssen for an ongoing research collaboration under which many of these results were first developed, and for extensive comments on this paper.  I also thank Pascal Vontobel for many helpful comments, and Moshe Schwartz for help with elementary graph theory.

\pagebreak

\end{document}